\documentclass[oldversion]{aa}
\usepackage{natbib}
\bibpunct{(}{)}{;}{a}{}{,}
\usepackage{graphicx}
\usepackage{txfonts}

\newcommand{\kms}{km~s$^{-1}$}

\begin{document}

   \title{Determination of the mass of the neutron star in
   \object{SMC~X$-$1}, \object{LMC~X$-$4} and \object{Cen~X$-$3} with
   VLT/UVES\thanks{Based on observations obtained at the European
   Southern Observatory at Paranal, Chile (ESO program 68.D-0568)}}
   \titlerunning{Determination of the mass of the neutron star in
   \object{SMC~X$-$1}, \object{LMC~X$-$4} and \object{Cen~X$-$3}}

   \author{
             A. van der Meer\inst{1},
             L. Kaper\inst{1,2},
	     M. H. van Kerkwijk\inst{3},
             M. H. M. Heemskerk\inst{1},
	     \and
	     E.P.J. van den Heuvel\inst{1,2}
          }

   \authorrunning{A. van der Meer et al.}

   \offprints{A. van der Meer}

   \institute{
                Astronomical Institute ``Anton Pannekoek'',
                University of Amsterdam, Kruislaan 403, NL-1098 SJ
                Amsterdam, Netherlands \\
                \email{ameer@science.uva.nl}
                \and
                Center for High-Energy Astrophysics, Kruislaan 403, NL-1098 SJ \\
                \and
                Department of Astronomy and Astrophysics, University
                of Toronto, 60 St George Street, Toronto, ON M5S 3H8,
                Canada \\
             }

   \date{Received ??, ??; accepted ??, ??}

   \abstract{We present the results of a spectroscopic monitoring
  campaign of the OB-star companions to the eclipsing X-ray pulsars
  \object{SMC~X$-$1}, \object{LMC~X$-$4} and
  \object{Cen~X$-$3}. High-resolution optical spectra obtained with
  UVES on the ESO \textit{Very Large Telescope} are used to determine
  the radial-velocity orbit of the OB (super)giants with high
  precision.  The excellent quality of the spectra provides the
  opportunity to measure the radial-velocity curve based on individual
  lines, and to study the effect of possible distortions of the line
  profiles due to e.g. X-ray heating on the derived radial-velocity
  amplitude. Several spectral lines show intrinsic variations with orbital
  phase. The magnitude of these variations depends on line strength,
  and thus provides a criterion to select lines that do not suffer
  from distortions. The undistorted lines show a larger
  radial-velocity amplitude than the distorted lines, consistent with
  model predictions.  Application of our line-selection criteria
  results in a mean radial-velocity amplitude $K_{\rm opt}$ of $20.2
  \pm 1.1$, $35.1 \pm 1.5$, and $27.5 \pm 2.3$~\kms\ ($1 \sigma$
  errors), for the OB companion to \object{SMC~X$-$1},
  \object{LMC~X$-$4} and \object{Cen~X$-$3}, respectively. Adding
  information on the projected rotational velocity of the OB companion
  (derived from our spectra), the duration of X-ray eclipse and
  orbital parameters of the X-ray pulsar (obtained from literature),
  we arrive at a neutron star mass of $1.06^{+0.11}_{-0.10}$,
  $1.25^{+0.11}_{-0.10}$ and $1.34^{+0.16}_{-0.14}$~M$_{\sun}$ for
  \object{SMC~X$-$1}, \object{LMC~X$-$4} and \object{Cen~X$-$3},
  respectively. The mass of \object{SMC~X$-$1} is near the minimum
  mass ($\sim$1~M$_{\sun}$) expected for a neutron star produced in a
  supernova. We discuss the implications of the measured mass
  distribution on the neutron-star formation mechanism, in relation to
  the evolutionary history of the massive binaries.

   }

   \maketitle


\section{Introduction}
\label{intro4}

  A neutron star is the compact remnant of a massive star (M $\gtrsim
  8$~M$_{\sun}$) with a central density that can be as high as 5 to 10
  times the density of an atomic nucleus. The global structure of a
  neutron star depends on the equation of state (EOS) under these
  extreme conditions, i.e. the relation between pressure and density
  in the neutron star interior (e.g. \citealt{lat04}). Given an EOS, a
  mass-radius relation for the neutron star and a corresponding
  maximum neutron-star mass can be derived. The ``stiffness'' of the
  EOS depends e.g. on how many bosons are present in matter of such a
  high density. As bosons do not contribute to the fermi pressure,
  their presence will tend to ``soften'' the EOS. For a soft EOS, the
  maximum neutron-star mass will be low (e.g. $<1.55$~M$_{\sun}$ for the EOS
  applied by \citet{bro94}); for a higher mass, the object would
  collapse into a black hole.

  The accurate measurement of neutron-star masses is therefore
  important for our understanding of the EOS of matter at
  supra-nuclear densities. Currently, the most massive neutron star in
  an X-ray binary is the X-ray pulsar \object{Vela~X$-$1}
  \citep{bar01,qua03} with a mass of $1.86 \pm 0.16$~M$_{\sun}$. The
  millisecond radio pulsar \object{J0751$+$1807} likely has an even
  higher mass: $2.1 \pm 0.2$~M$_{\sun}$ \citep{nic05}. Both results
  are in favor of a stiff EOS (see also \citealt{sri01}). Neutron
  stars also have a minimum mass limit. The minimum stable
  neutron-star mass is about 0.1~M$_{\sun}$, although a more realistic
  minimum stems from a neutron star's origin in a
  supernova. Lepton-rich proto neutron stars are unbound if their
  masses are less than about 1~M$_{\sun}$ \citep{lat04,hae02}.

  Another issue is the neutron-star mass distribution: the detailed
  supernova mass ejection mechanism accompanying the formation of the
  neutron star is not understood, but it is likely that the many
  neutrinos that are produced during the formation of the (proto-)
  neutron star in the centre of the collapsing star play an important
  role (e.g. \citealt{burr00}). \citet{tim96} present model
  calculations from which they conclude that Type~II supernovae
  (massive, single stars) will give a bimodal neutron-star mass
  distribution, with peaks at 1.28 and 1.73 M$_{\sun}$, while Type~Ib
  supernovae (such as produced by stars in binaries, which are
  stripped of their envelopes) will produce neutron stars within a
  small range around 1.32~M$_{\sun}$. Despite the fact that it is in a
  binary, the massive neutron star in \object{Vela~X$-$1} may belong
  to the second peak in this mass distribution.
  \begin{table*}[!ht]
  \caption[]{Orbital parameters of \object{SMC~X$-$1},
  \object{LMC~X$-$4} and \object{Cen~X$-$3} obtained from
  \citet{woj98}, \citet{lev00} and \citet{nag92},
  respectively. $T_{0}$ is the mid-eclipse time and corresponds to
  orbital phase $\phi = 0.0$. The eccentricity is obtained from
  \citet{bil97} and references therein. All the reported errors are
  $1\sigma$ values unless stated otherwise. For \object{SMC~X$-$1} the
  range of semi-eclipse angle $\theta_{\rm e}$ includes observations
  obtained by \citet{pri76}, \citet{bon81} and \citet{sch72a}; for
  \object{LMC~X$-$4} those of \citet{li78}, \citet{whi78} and
  \citet{pie85}; for \object{Cen~X$-$3} we list the result of
  \citet{cla88}.}
  \label{literature_values4}
  \begin{center}
  \begin{tabular}{p{3.0cm}p{3.0cm}p{3.0cm}p{3.0cm}} 
  \hline \hline
  & \object{SMC~X$-$1} & \object{LMC~X$-$4} & \object{Cen~X$-$3} \\
  \hline
  &&&\\[-0.3cm]
  $T_{0}$ (MJD)                                 & $42836.18278(20)$               & $51110.86579(10)$        & $40958.35(1)$                \\
  $P_{\rm orb}$ (days)                          & $3.89229090(43)$                & $1.40839776(26)$         & $2.08713845(5)$              \\
  $\dot{P}_{\rm orb}$/$P_{\rm orb}$ (yr$^{-1}$) & $-3.353(14) \times 10^{-6}$     & $-9.8(7) \times 10^{-7}$ & $-1.738(4) \times 10^{-6}$   \\
  $P_{\rm spin}$ (sec)                          & $0.708$                         & $13.5$                   & $4.82$                       \\
  $a_{\rm X} \sin{i}$ (lt-sec)                  & $53.4876(4)$                    & $26.343(16)$             & $39.56(7)$                   \\
  $e$                                           & $< 4 \times 10^{-5} (2 \sigma)$ & $0.006(2)$               & $<1.6 \times 10^{-3} (90\%)$ \\
  $\theta_{\rm e}$ (deg)                        & $26-30.5$                       & $25-29$                  & $32.9 \pm 1.4$               \\ 
  OB companion                                 & B$0$~Ib                         & O$8$~III                 & O$6.5$~II-III                \\
  \hline
  \end{tabular}
  \end{center}
  \end{table*}

  Neutron stars are detected either as radio pulsars, single or in a
  binary with a white dwarf or neutron star companion, or as X-ray
  sources in binaries with a (normal) low-mass (LMXB)
  or a high-mass companion star (HMXB). Presently, all accurate mass
  determinations have been for neutron stars that were almost
  certainly formed in Type~Ib supernovae and that have accreted little
  since. Exceptions are \object{J1909$-$3744}, a pulsar ($+$ white
  dwarf) with a mass of $1.438 \pm 0.024$~M$_{\sun}$ \citep{jac05},
  and the massive neutron star in \object{J0751$+$1807}, which may
  have originated from an LMXB. The most accurate masses have been
  derived for the binary radio pulsars. Until recently, all of these
  were consistent with a small mass range near 1.35~M$_{\sun}$
  \citep{tho99}.

  We focus here on the initially most massive systems, which consist
  of a massive OB supergiant and a neutron star or a black hole
  \citep{kap01,kap05}. The main motivation to concentrate on these
  systems is that they are the most likely hosts of massive neutron
  stars. About a dozen of these systems are known\footnote{Recently,
  several new hard X-ray sources have been detected with INTEGRAL that
  show the characteristics of a heavily obscured HMXB with an
  OB-supergiant companion, called supergiant fast X-ray transients
  \citep{neg06,lut05}}; five of them contain an eclipsing X-ray
  pulsar. The masses of all but one (\object{Vela~X$-$1}) are
  consistent (within their errors) with a value of about
  $1.4$~M$_{\sun}$. However, most spectroscopic observations used for
  these mass determinations were carried out more than 20 years ago,
  before the advent of sensitive CCD detectors and 8m-class
  telescopes, which allow high-resolution spectroscopy of the optical
  companions. The uncertainties in the earlier radial velocity
  measurements (see \citealt{ker95b}) are too large to measure a
  significant spread in mass among these neutron stars, if present.

  In this paper we present new, more accurate determinations of the
  mass of the neutron star in three of these systems,
  i.e. \object{SMC~X$-$1}, \object{LMC~X$-$4}, and \object{Cen~X$-$3}
  using the high-resolution Ultraviolet and Visual Echelle
  Spectrograph (UVES) on the ESO \textit{Very Large Telescope}
  (VLT). These systems are in a phase of Roche-lobe overflow
  \citep{sav78,sav83}, have well determined, circular orbits ($P_{\rm
  orb}$ of a few days), and an optical counterpart of $V \simeq
  14$~mag, i.e. well within reach of VLT/UVES.

  In Sect.~\ref{hmxbs4} we introduce the three HMXBs. In
  Sect.~\ref{observations4} we describe the acquired observations and
  data reduction procedure. In Sect.~\ref{spectra4} we present the
  spectral analysis and the resulting radial-velocity curves. In
  Sect.~\ref{nsmasses4} we evaluate the measured radial-velocity
  amplitudes and derive the mass of the neutron star in these three
  systems. In Sect.~\ref{conclusions4} we summarise our conclusions and
  in Sect.~\ref{discussion4} we compare these to the predictions of
  supernova models.


\section{Eclipsing high-mass X-ray binaries; review of earlier work}
\label{hmxbs4}

  Five high-mass X-ray binaries are known to host an eclipsing X-ray
  pulsar: \object{Vela~X$-$1}, \object{4U~1538$-$52},
  \object{SMC~X$-$1}, \object{LMC~X$-$4} and \object{Cen~X$-$3}. The
  eclipse provides an important constraint on the orbital inclination
  $i$, an essential parameter for the mass determination. For the
  eclipsing X-ray source \object{4U~1700$-$37} with the O6.5~Iaf+
  companion \object{HD~153919} \citep{jon73,mas76} no X-ray pulsations
  have been detected, although the compact object most likely is a
  neutron star (\citealt{rey99}; \citealt{mee05}). The absence of
  X-ray pulsations prohibits the accurate determination of the orbital
  parameters of the neutron star, and thus its mass.
  \begin{table*}[!ht]
  \caption[]{Observing log of the three observed systems. For each
  observation the Modified Julian Date (MJD) corresponding to the
  mid-exposure time is given, the orbital phase ($\phi$) and the
  signal-to-noise ratio (S/N) per resolution element of the three
  different CCDs centered at 4270~{\AA} for the blue CCD, 5150~{\AA}
  for the red1 CCD and 6100~{\AA} for the red2 CCD.}
  \label{obs_log4}
  \begin{center}
  \begin{tabular}{ p{1.7cm} p{0.5cm} *{3}{p{0.1cm}}| p{1.7cm} p{0.5cm} *{3}{p{0.1cm}}| p{1.7cm} p{0.5cm} *{3}{p{0.1cm}}}
  \hline \hline
  \multicolumn{5}{c|}{\object{SMC~X$-$1}}
  & \multicolumn{5}{|c|}{\object{LMC~X$-$4}}
  & \multicolumn{5}{|c}{\object{Cen~X$-$3}}
  \\
  \hline
    \multicolumn{1}{l}{MJD}
  & \multicolumn{1}{l}{orbital}
  & \multicolumn{3}{c|}{S/N}
  & \multicolumn{1}{|l}{MJD}
  & \multicolumn{1}{l}{orbital}
  & \multicolumn{3}{c|}{S/N}
  & \multicolumn{1}{|l}{MJD}
  & \multicolumn{1}{l}{orbital}
  & \multicolumn{3}{c}{S/N}
  \\
    \multicolumn{1}{l}{(days)}
  & \multicolumn{1}{l}{phase($\phi$)}
  & \multicolumn{1}{l}{blue\hspace*{-0.3cm}}
  & \multicolumn{1}{l}{red1\hspace*{-0.3cm}}
  & \multicolumn{1}{l|}{red2}
  & \multicolumn{1}{|l}{(days)}
  & \multicolumn{1}{l}{phase($\phi$)}
  & \multicolumn{1}{l}{blue\hspace*{-0.3cm}}
  & \multicolumn{1}{l}{red1\hspace*{-0.3cm}}
  & \multicolumn{1}{l|}{red2}
  & \multicolumn{1}{|l}{(days)}
  & \multicolumn{1}{l}{phase($\phi$)}
  & \multicolumn{1}{l}{blue\hspace*{-0.3cm}}
  & \multicolumn{1}{l}{red1\hspace*{-0.3cm}}
  & \multicolumn{1}{l}{red2}
  \\
  \hline
  & & & & & & & & & & & & & &  \\[-0.3cm]
  52187.154 & 0.537 & 58 & 75 & 69 & 52214.187 & 0.388 & 54 & 69 & 58 & 52271.300 & 0.462 & 33 & 67 & 77 \\
  52214.090 & 0.458 & 60 & 80 & 71 & 52224.249 & 0.533 & 43 & 58 & 50 & 52287.229 & 0.095 & 22 & 50 & 64 \\
  52224.224 & 0.062 & 49 & 69 & 62 & 52225.255 & 0.247 & 49 & 66 & 57 & 52292.273 & 0.512 & 27 & 57 & 68 \\
  52225.211 & 0.315 & 58 & 78 & 68 & 52226.212 & 0.927 & 48 & 67 & 56 & 52298.169 & 0.337 & 32 & 63 & 75 \\
  52242.059 & 0.644 & 41 & 61 & 54 & 52237.314 & 0.809 & 54 & 72 & 63 & 52309.323 & 0.681 & 35 & 76 & 81 \\
  52243.087 & 0.908 & 66 & 77 & 70 & 52242.282 & 0.337 & 55 & 71 & 61 & 52313.150 & 0.515 & 37 & 74 & 78 \\
  52244.168 & 0.186 & 67 & 81 & 74 & 52243.113 & 0.927 & 54 & 62 & 55 & 52321.163 & 0.354 & 29 & 64 & 72 \\
  52247.185 & 0.961 & 43 & 58 & 51 & 52244.193 & 0.694 & 59 & 67 & 55 & 52322.158 & 0.831 & 31 & 65 & 73 \\
  52256.112 & 0.255 & 53 & 66 & 59 & 52245.213 & 0.418 & 67 & 73 & 65 & 52326.186 & 0.761 & 38 & 77 & 82 \\
  52258.094 & 0.764 & 58 & 69 & 62 & 52246.085 & 0.037 & 44 & 53 & 45 & 52327.277 & 0.284 & 37 & 77 & 85 \\
  52270.052 & 0.837 & 61 & 74 & 68 & 52256.086 & 0.138 & 32 & 40 & 35 & 52328.241 & 0.746 & 41 & 82 & 88 \\
  52271.059 & 0.095 & 52 & 65 & 60 & 52259.094 & 0.274 & 49 & 59 & 50 & 52329.260 & 0.234 & 38 & 76 & 86 \\
  52285.112 & 0.706 & 47 & 61 & 56 & 52261.163 & 0.743 & 58 & 71 & 61 &             &         &    &    &    \\
  \hline
  \end{tabular}
  \end{center}
  \end{table*}

  \citet{ker95b} present an analysis of the neutron-star mass
  determinations for these systems hosting an X-ray pulsar and
  conclude that the accuracy of the (then) available observations does
  not allow to discriminate between one ``canonical'' neutron-star
  mass or a mass distribution. Recent analyses of the radial-velocity
  curve of the wind-fed system \object{Vela~X$-$1} \citep{bar01,qua03}
  with its B0.5~Ib companion \citep{hil72,vid73} have shown that the
  neutron star in this system has a mass of $1.86 \pm
  0.16$~M$_{\sun}$. Such a high neutron-star mass provides an
  important constraint on the EOS at supra-nuclear density.

  Since the work of \citet{rey92}, included in the analysis by
  \citet{ker95b}, no new optical spectroscopy of the B0~supergiant
  companion (\object{QV~Nor}) of \object{4U~1538$-$52} has been
  reported in literature. \citet{ker95b} list
  $1.06^{+0.41}_{-0.34}$~M$_{\sun}$ for the mass of
  \object{4U~1538$-$52}.

  We have obtained VLT/UVES spectra of the three Roche-lobe overflow
  systems \object{SMC~X$-$1}, \object{LMC~X$-$4} and
  \object{Cen~X$-$3}. The orbital parameters
  (Table~\ref{literature_values4}) of their X-ray pulsars are
  accurately known, based on X-ray pulse time delay measurements. The
  X-ray pulsars in these systems have short spin periods (seconds)
  compared to those in wind-fed systems (minutes), as the mass- and
  angular-momentum accretion rate in Roche-lobe overflow systems is
  much higher than in wind-fed systems. The photometric light curves
  indicate that in all three systems an accretion disc is present
  \citep{tje86,hee89}. The X-ray eclipse duration is best measured in
  hard X-rays, since at lower energies the eclipses are systematically
  longer due to soft X-ray absorption by the stellar wind of the
  OB~companion. The eclipse duration can thus be used to determine the
  radius of the OB~companion.

\subsection{\object{SMC~X$-$1}}

  The B0~supergiant \object{Sk~160} ($V = 13.3$~mag) is the companion
  to the eclipsing X-ray pulsar \object{SMC~X$-$1}
  \citep{sch72a,lil73}, located in the ``wing'' of the Small
  Magellanic Cloud at a distance of 60.6~kpc \citep{hil05}. The spin
  period of the pulsar is 0.71~s and the orbital period is 3.89~d,
  which is decaying on a timescale of $3 \times 10^{6}$~yr due to
  tidal interaction \citep{lev93}. A super-orbital, though not strictly
  periodic variation of $\sim 60$~d is present in the system, most
  likely due to a precessing tilted accretion disc
  \citep{woj98,cla03}.

  The most recent determination of the radial-velocity orbit of
  \object{Sk~160} has been performed by \citet{bak05}.  Optical
  spectra covering the wavelength range 4300--5100~{\AA} were obtained
  with the grating spectrograph on the 1.9 metre Radcliff telescope at
  the Sutherland Observatory, with a resolving power $R \sim
  4000$. The majority of the 56 usable spectra were secured during one
  week of observations in September 2000. Based on a cross-correlation
  analysis similar to the one used by \citet{rey93}, a radial-velocity
  amplitude of $K_{\rm opt} = 18.0 \pm 1.8$~\kms\ was measured, which
  becomes $21.8 \pm 1.8$~\kms\ when taking the effects of X-ray
  heating into account; the rest-frame ($\gamma$) velocity is
  174~\kms. To simulate the effects of X-ray heating, a model is used
  that generates velocity corrections based on contributions from
  different elements of the projected stellar disc. The models do not
  take into account the presence of an accretion disc, which may well
  reduce the effect of X-ray heating \citep{ker95b}. The results of
  \citet{bak05} are significantly different from the results obtained
  by \citet{rey93} who arrive at $K_{\rm opt} = 27.5 \pm 1.9$~\kms\
  following a similar procedure. According to \citet{bak05}, this
  discrepancy could be due to the limited phase coverage of the
  dataset of \citet{rey93} and to the fact that \citet{rey93} assume a
  significantly higher value for $L_{X}$ when determining the
  non-Keplerian corrections. The latter would, however, not explain
  the difference in $K_{\rm opt}$ \textit{before} applying the X-ray
  heating corrections. \citet{bak05} derive lower (edge-on system) and
  upper limits (Roche-lobe filling system) to the mass of
  \object{SMC~X$-$1} of $0.91 \pm 0.08$~M$_{\sun}$ and $1.21 \pm
  0.10$~M$_{\sun}$, respectively. The mass of the optical companion
  is around $16.7$~M$_{\sun}$ in both cases.

\subsection{\object{LMC~X$-$4}}

  After the first detection of \object{LMC~X$-$4} by the \textit{Uhuru}
  satellite \citep{gia72}, the binary nature of its optical
  counterpart was confirmed by \citet{che77}. The $V = 14.0$~mag
  O8~III companion (\citealt{san76}; Kaper~et~al., to be submitted) is
  in a 1.41~d orbit \citep{li78,whi78}, which is decaying on a
  timescale of $\sim 500\,000$~yr \citep{lev00}. The optical light curve
  shows ellipsoidal variations and a super-orbital period of $\sim
  30$~d due to a precessing accretion disc \citep{hee89}. The X-ray
  light curve includes regular eclipses as well as a pronounced flux
  modulation of a factor $\sim 60$ with a period of 30.5~d
  \citep{lan81}. This long-term variation is attributed to the
  precessing accretion disc. \citet{kel83} discovered the 13.5~s X-ray
  pulsations of \object{LMC~X$-$4}.

  \citet{che77} reported on photographic spectra obtained with the
  1.5m~ESO telescope from which they derived radial-velocity
  variations with an amplitude of $475 \pm 25$~\kms\ for the
  $\ion{He}{ii}$ 4686~{\AA} line, and a phase dependence suggesting an
  origin near the X-ray source. \citet{hut78} collected 18
  spectrograms using the Cassegrain image-tube spectrograph of the
  CTIO 4m telescope in November 1977, with an effective spectral
  resolution of $R \sim 3000$. For the hydrogen lines
  ($\ion{H\beta}{}$ to $\ion{H9}{}$, and an empirically determined
  correction to the blended $\ion{H\delta}{}$ line) they derive a
  radial-velocity amplitude of $K_{\rm opt} = 50 \pm 5$~\kms; for the
  $\ion{He}{i}$ lines $K_{\rm opt} = 60 \pm 9$~\kms\ is measured. For
  the $\ion{He}{ii}$ 4686~{\AA} line \citet{hut78} derive $K_{\rm opt} =
  498 \pm 14$~\kms, with a phase difference compared to the
  $\ion{H}{}$ and $\ion{He}{i}$ absorption lines of 0.79~$P_{\rm
  orb}$, consistent with \citet{che77}.
  \citet{kel83} combined the radial-velocity data of the hydrogen
  absorption lines presented by \citet{hut78} with measurements by
  \citet{pet82} and arrive at $K_{\rm opt} = 37.9 \pm
  2.4$~\kms. \citet{ker95b} use $K_{\rm opt} = 38 \pm 5$~\kms\ and obtain
  M$_{X} = 1.47^{+0.44}_{-0.39}$~M$_{\sun}$ and M$_{\rm opt} =
  15.8^{+2.3}_{-2.0}$~M$_{\sun}$.
  \begin{figure*}[!ht]
  \centering
  \includegraphics[width=17.5cm]{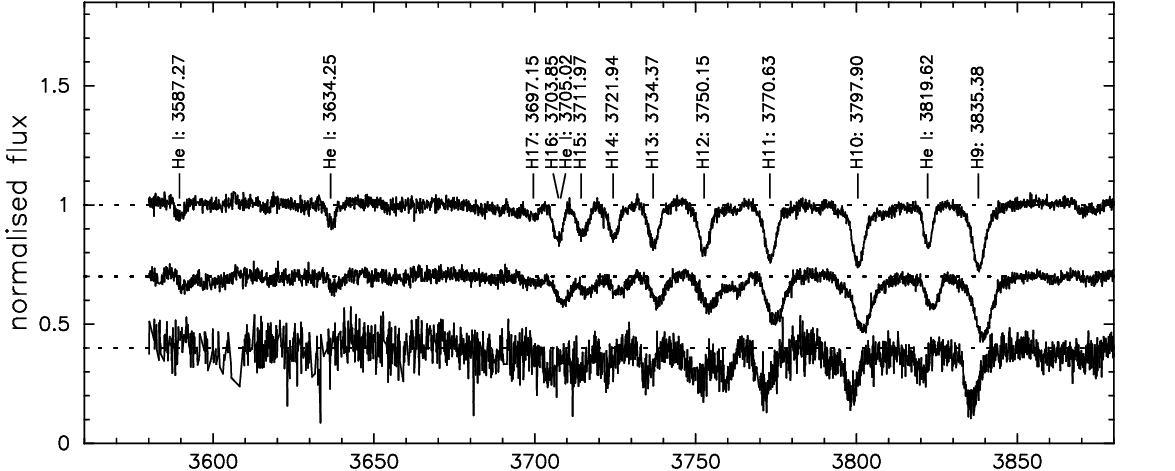}

  \vspace{0.8cm}
  \includegraphics[width=17.5cm]{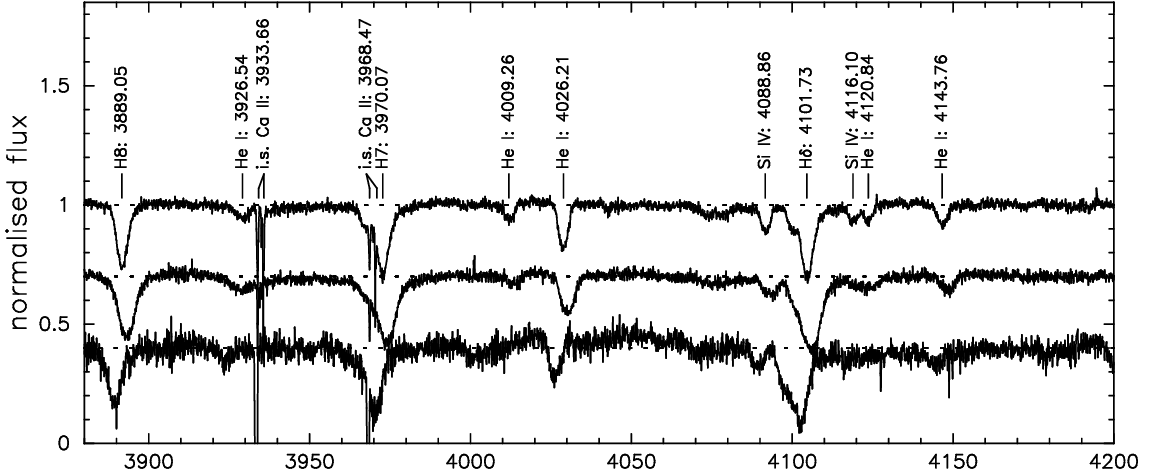}

  \vspace{0.8cm}
  \includegraphics[width=17.5cm]{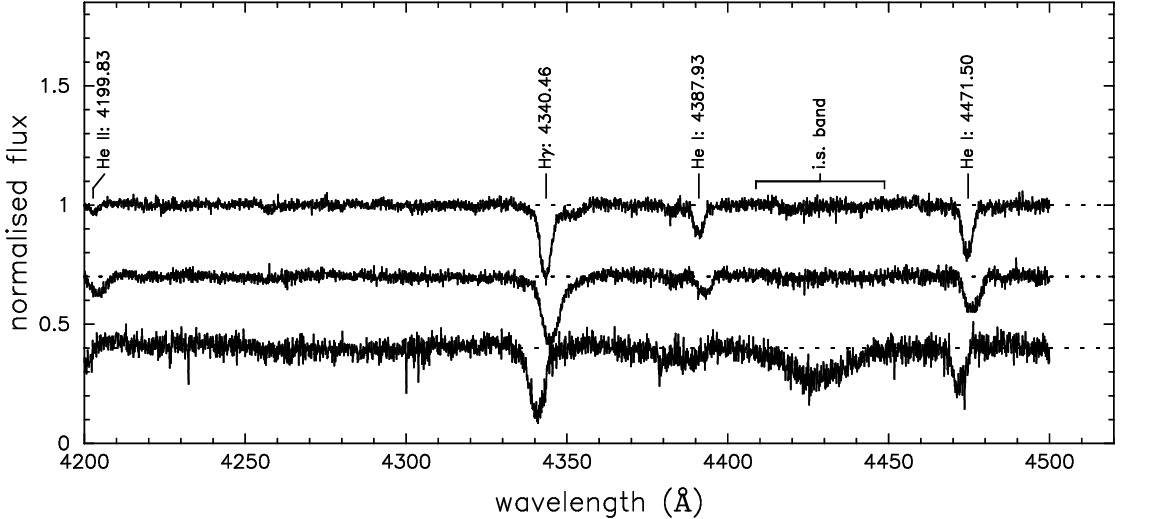}
  \caption{Normalised spectra in the wavelength range 3580--4500~{\AA}
  of \object{SMC~X$-$1} at orbital phase $\phi = 0.06$ (top spectrum),
  \object{LMC~X$-$4} at orbital phase $\phi = 0.04$ (middle spectrum) and
  \object{Cen~X$-$3} at orbital phase $\phi = 0.09$ (bottom spectrum),
  respectively. The line identifications are shown above the spectrum
  of \object{SMC~X$-$1}. Note that the reddening of \object{Cen~X$-$3}
  ($E(B-V) \sim 1.4$) clearly affects the S/N in this wavelength
  range.}
  \label{spectra_blue4}
  \end{figure*}
  \begin{figure*}[!ht]
  \centering
  \includegraphics[width=17.5cm]{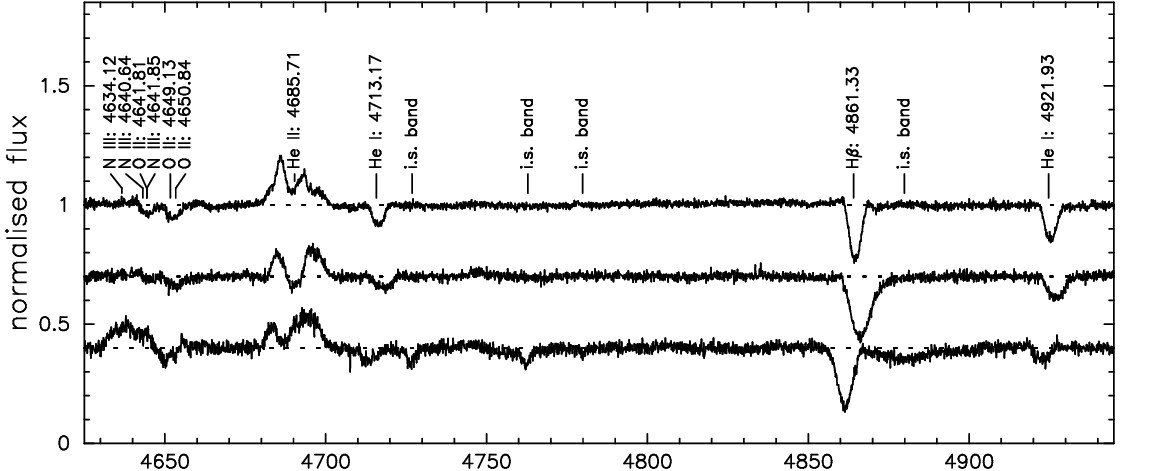}

  \vspace{0.8cm}
  \includegraphics[width=17.5cm]{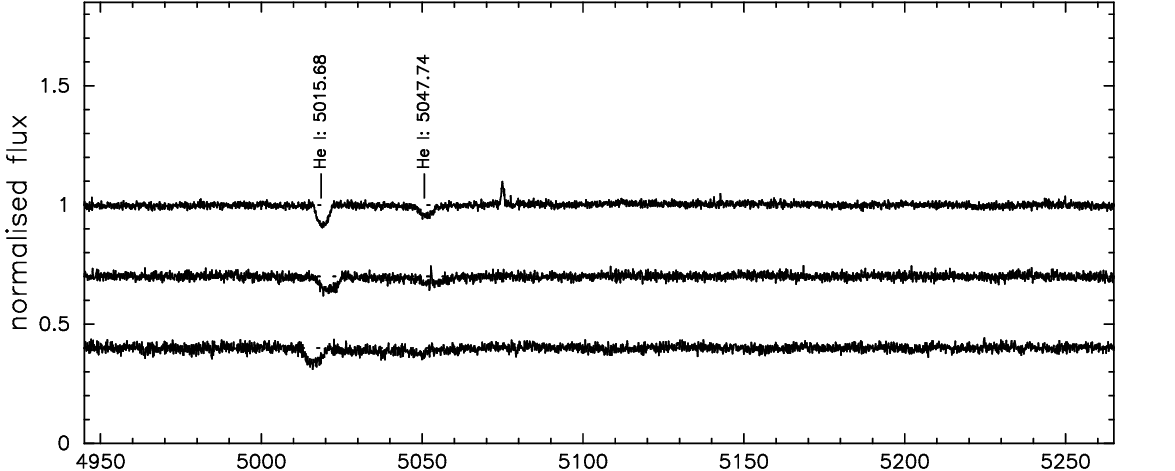}

  \vspace{0.8cm}
  \includegraphics[width=17.5cm]{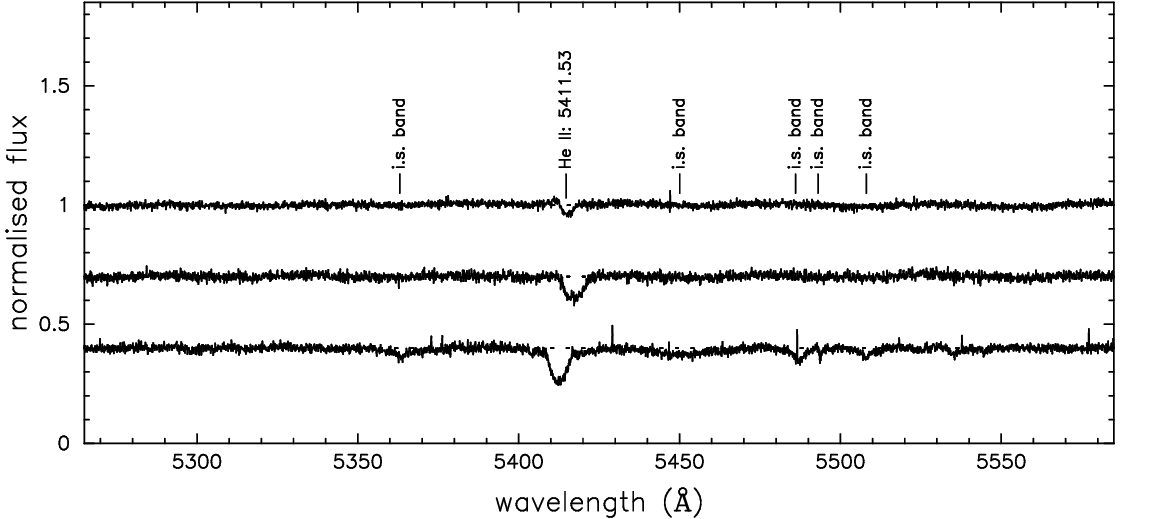}
  \caption{Normalised spectra in the wavelength range 4625--5585~{\AA}
  of \object{SMC~X$-$1} at orbital phase $\phi = 0.06$ (top spectrum),
  \object{LMC~X$-$4} at orbital phase $\phi = 0.04$ (middle spectrum) and
  \object{Cen~X$-$3} at orbital phase $\phi = 0.09$ (bottom spectrum),
  respectively. The line identifications are shown above the spectrum
  of \object{SMC~X$-$1}.}
  \label{spectra_red14}
  \end{figure*}
  \begin{figure*}[!ht]
  \centering
  \includegraphics[width=17.5cm]{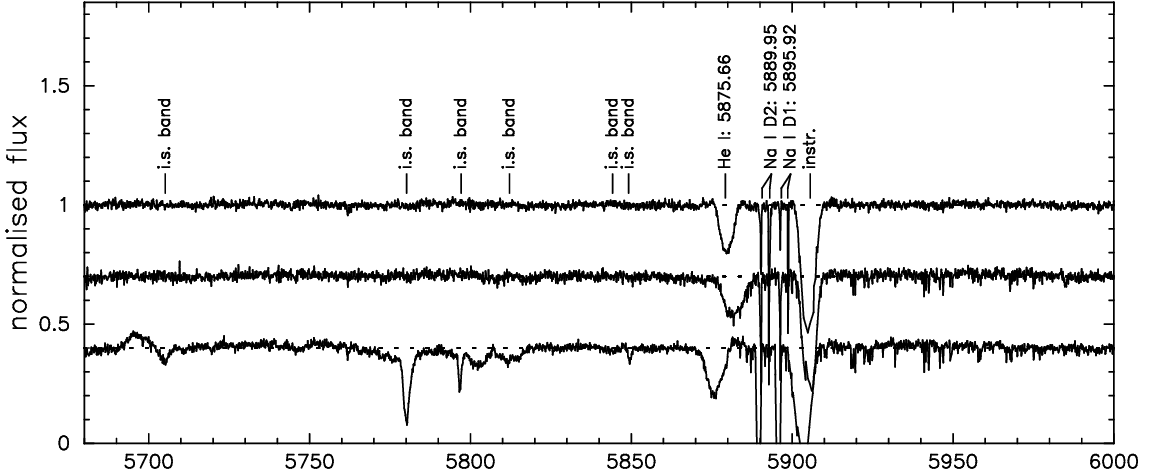}

  \vspace{0.8cm}
  \includegraphics[width=17.5cm]{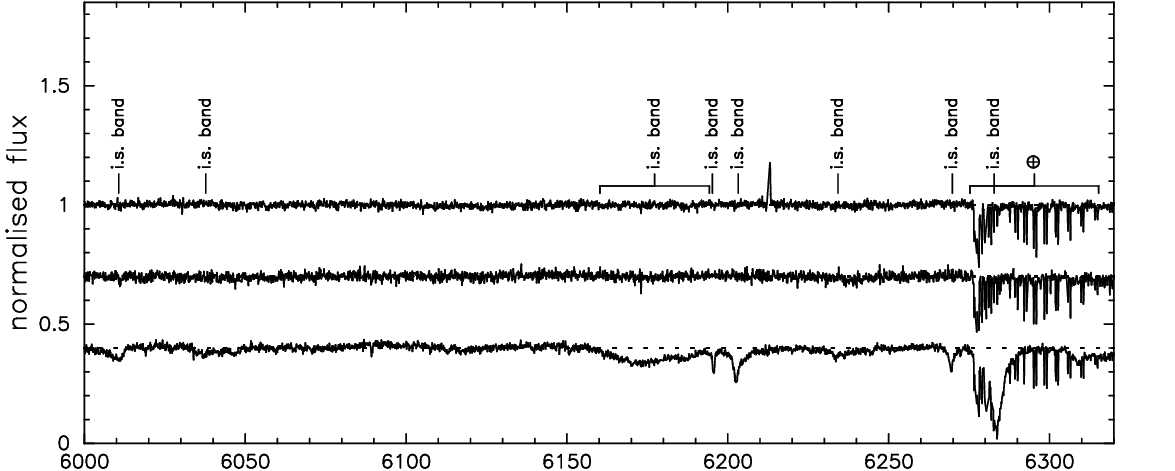}

  \vspace{0.8cm}
  \includegraphics[width=17.5cm]{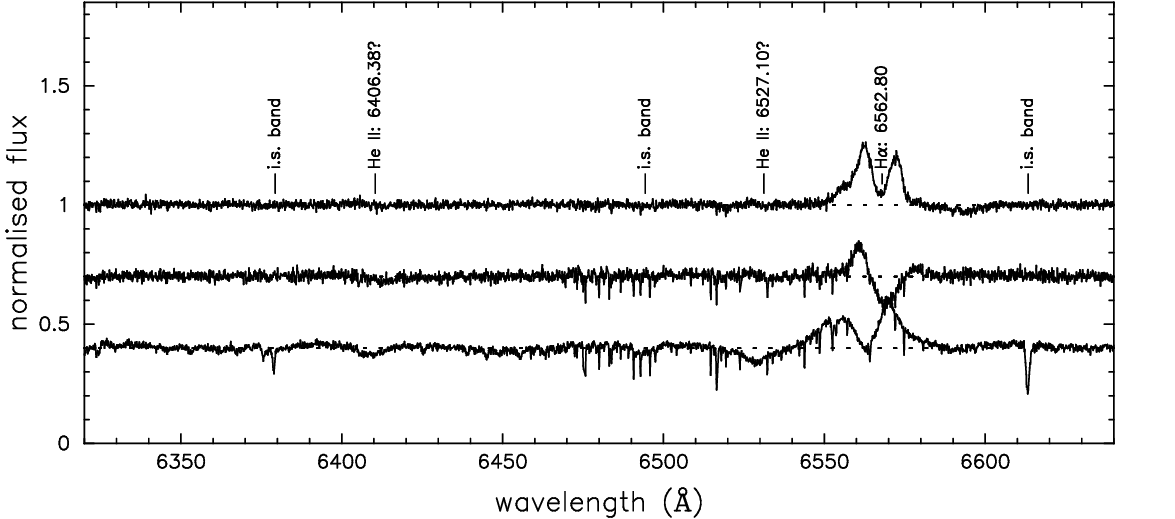}
  \caption{Normalised spectra in the wavelength range 5680--6640~{\AA}
  of \object{SMC~X$-$1} at orbital phase $\phi = 0.06$ (top spectrum),
  \object{LMC~X$-$4} at orbital phase $\phi = 0.04$ (middle spectrum) and
  \object{Cen~X$-$3} at orbital phase $\phi = 0.09$ (bottom spectrum),
  respectively. The line identifications are shown above the spectrum
  of \object{SMC~X$-$1}.}
  \label{spectra_red24}
  \end{figure*}

\subsection{\object{Cen~X$-$3}}

  \object{Cen~X$-$3} was discovered by \citet{cho67} and became the
  first detected binary X-ray pulsar \citep{gia71,sch72b}. The $V =
  13.3$~mag optical counterpart \object{V779~Cen} was identified by
  \citet{krz74}, an O6-7 II-III star \citep{ash99} in a 2.09~d
  circular orbit with the 4.84~s X-ray pulsar. The optical light curve
  indicates the likely presence of an accretion disc, but no strong
  evidence is found for X-ray heating \citep{tje86}. The X-ray
  light curve includes episodes of high and low X-ray flux with a
  characteristic timescale of 120--165~d \citep{pri83,pau05}.
  \begin{figure}[!t]
  \centering
  \includegraphics[width=7.2cm]{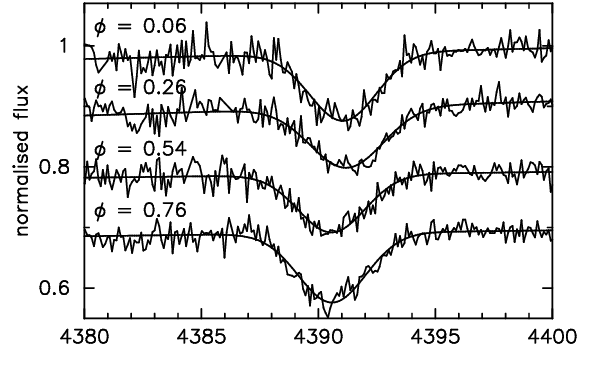}

  \vspace{-0.2cm}
  \includegraphics[width=7.2cm]{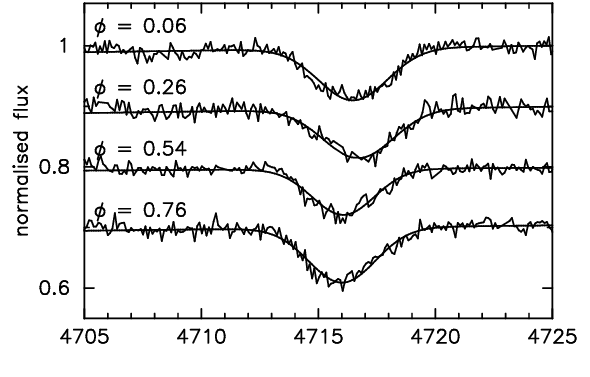}

  \vspace{-0.2cm}
  \includegraphics[width=7.2cm]{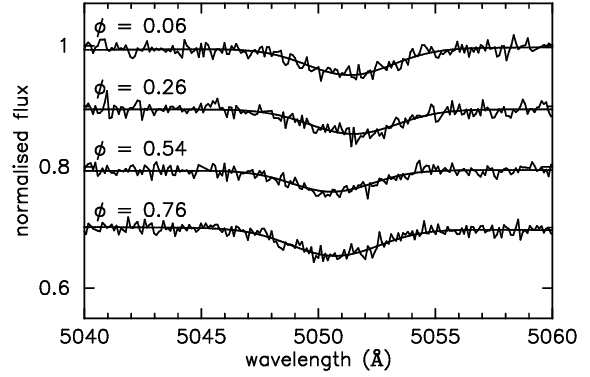}
  \caption{Example of gaussian fits for the \ion{He}{i} line at
  4387.93~{\AA} (upper panel), the \ion{He}{i} line at 4713.17~{\AA}
  (middle panel) and the \ion{He}{i} line at 5047.74~{\AA} (lower panel)
  in the spectrum of the B supergiant companion to \object{SMC~X$-$1}. The
  spectra and the gaussian fits are shown at four different quadratures. 
  To better visualise the
  fits the spectra are smoothed. Note that the gaussian profile fits
  represent the rotationally broadened line profiles very well. Furthermore,
  the orbital motion of the B supergiant companion is easily observed.}
  \label{gaussfits4}
  \end{figure}

  Based on photographic spectra, \citet{hut79} derive $K_{\rm opt} =
  24 \pm 6$~\kms, confirmed by \citet{asl82} who also report $K_{\rm
  opt} = 24 \pm 6$~\kms. The most recent radial-velocity measurements
  of \object{V779~Cen} are presented by \citet{ash99} who determine
  two very different values of $K_{\rm opt}$ based on two datasets
  obtained with the 4m \textit{Anglo-Australian Telescope} and the RGO
  spectrograph. The wavelength range of these spectra is
  4300--4700~{\AA}; the spectral resolution $R \sim
  3000$. \citet{ash99} discard the results of the first dataset and
  arrive at $K_{\rm opt} = 24.4 \pm 4.1$~\kms. The resulting
  neutron-star mass is M$_{X} = 1.21 \pm 0.21$~M$_{\sun}$ and the mass
  of the O-type companion M$_{\rm opt} = 20.5 \pm 0.7$~M$_{\sun}$.
  \begin{figure}[!t]
  \centering
  \includegraphics[width=7.2cm]{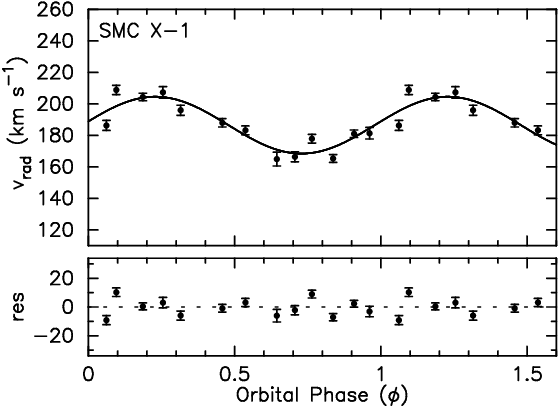}

  \vspace{0.2cm}
  \includegraphics[width=7.2cm]{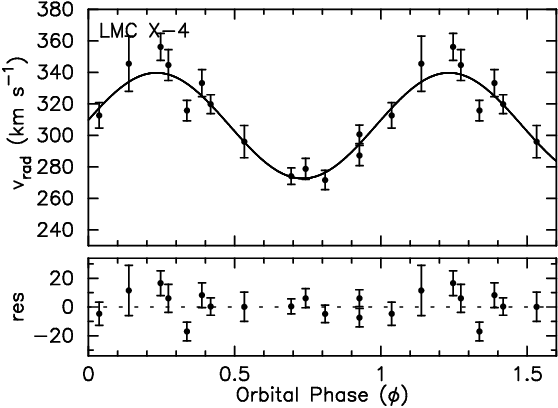}

  \vspace{0.2cm}
  \includegraphics[width=7.2cm]{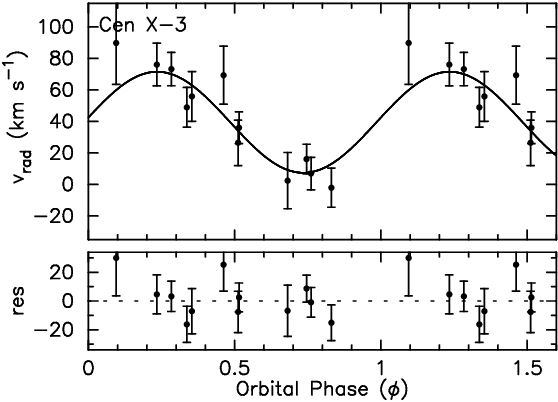}
  \caption{Example of radial-velocity curves obtained from the \ion{H13}{} line
  at 3734.37~{\AA}. Some datapoints are shown twice to better visualise trends
  with orbital phase. The error bars indicate 1 $\sigma$ errors. The
  lower panels show the residuals of the fit.}
  \label{radvels_oneline4}
  \end{figure}


\section{Observations}
\label{observations4}

  We have obtained high-resolution ($R \sim 40\,000$) spectra of the
  three systems with UVES \citep{dek00} on the VLT in service mode in
  the period October 2001 to March 2002 at Paranal, Chile. The total
  exposure time was 17.6~h spread over 13 exposures of 1400~s of
  \object{SMC~X$-$1}, 13 exposures of 2000~s of \object{LMC~X$-$4} and
  12 exposures of 1600~s of \object{Cen~X$-$3}. The instrument was
  used with standard setting ``390+564'' and a slit width of
  1.0\arcsec. This yields a wavelength range of 3580--4500~{\AA} in
  the blue arm and 4625--5585~{\AA} and 5680--6640~{\AA} in the red
  arm. To determine the orbital phase of the systems we used the
  ephemeris of \citet{woj98}, \citet{lev00} and \citet{nag92} for
  \object{SMC~X$-$1}, \object{LMC~X$-$4} and \object{Cen~X$-$3},
  respectively (see Table~\ref{literature_values4}). The log of
  observations is listed in Table~\ref{obs_log4}.

  In order to reduce the data to normalised spectra we used the UVES
  pipeline (version 1.2.0) and the ESO reduction package MIDAS
  (version 03SEPpl1.1). All raw echelle frames were bias and flatfield
  corrected. Subsequently the different orders were extracted using
  the optimal extraction routine available within the UVES pipeline.
  High signal-to-noise ($S/N \ga 50$) spectra extracted with this
  routine show a ripple effect. In our spectra this effect is
  only marginally present in some spectra of \object{SMC~X$-$1} in the
  wavelength range 5100--5585~{\AA}. This range contains the
  $\ion{He}{ii}$ 5411.53~{\AA} line, but we do not use it for the
  determination of the radial-velocity amplitude of the system
  (Sect.~\ref{line_selection4}). 
  \begin{figure*}[!ht] 
  \begin{center}
  \includegraphics[width=5.5cm]{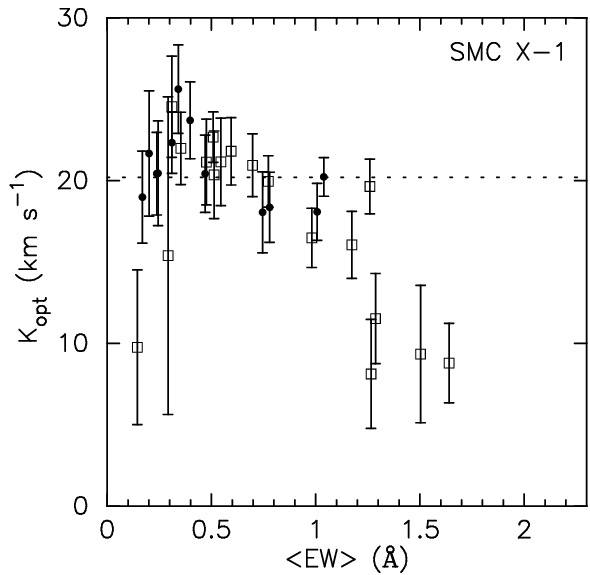}\hspace{-0.3cm}
  \includegraphics[width=5.5cm]{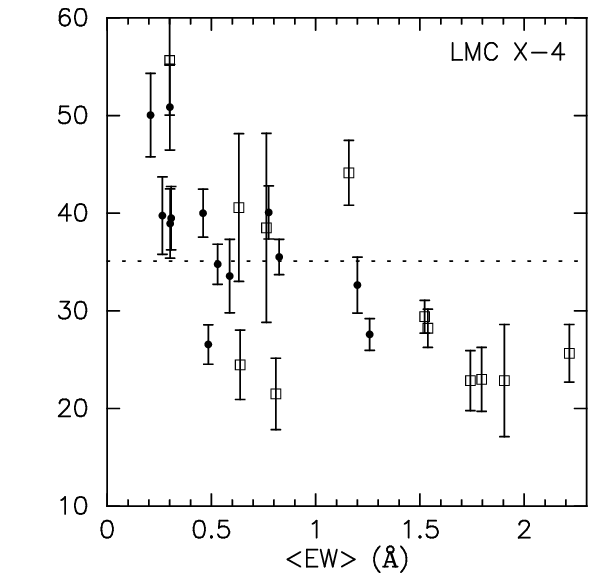}\hspace{-0.3cm}
  \includegraphics[width=5.5cm]{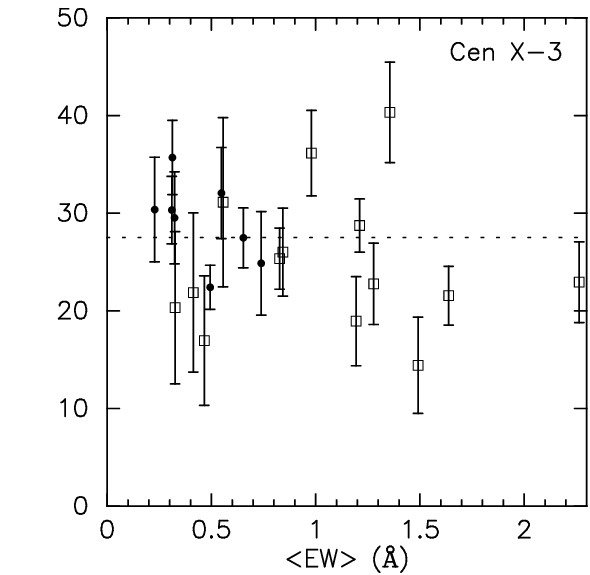} 

  \includegraphics[width=5.5cm]{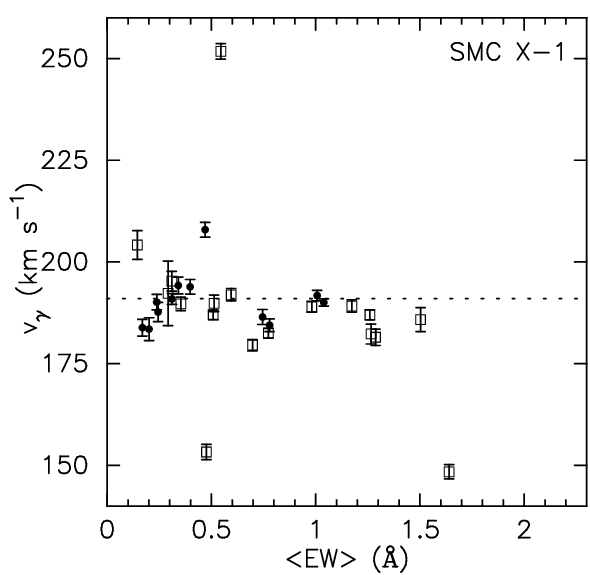}\hspace{-0.3cm}
  \includegraphics[width=5.5cm]{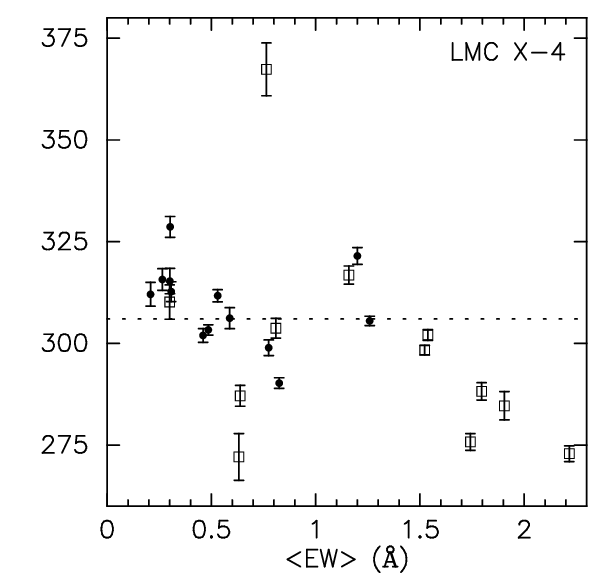}\hspace{-0.3cm}
  \includegraphics[width=5.5cm]{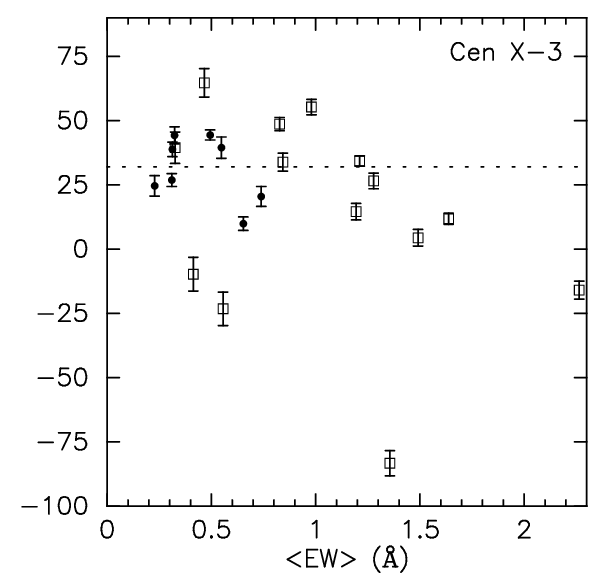}

  \includegraphics[width=5.5cm]{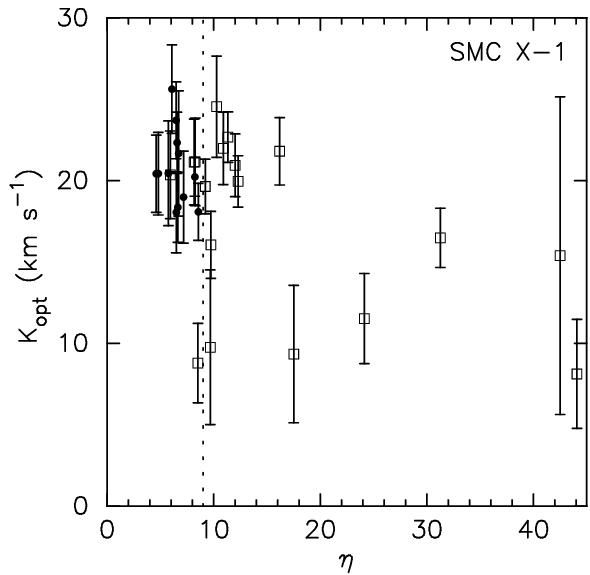}\hspace{-0.3cm}
  \includegraphics[width=5.5cm]{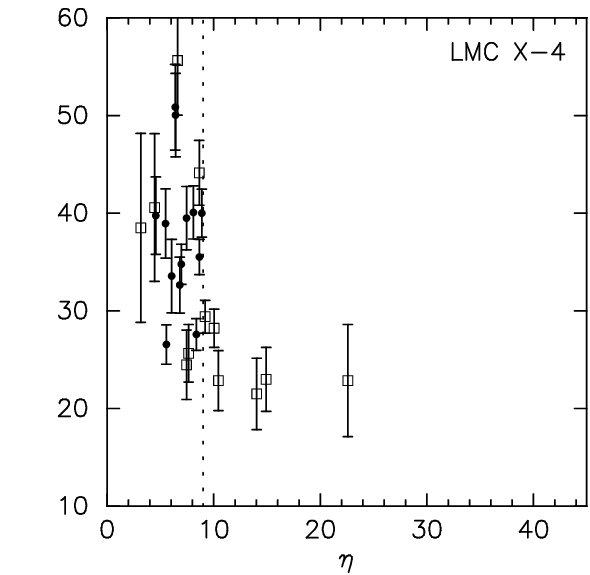}\hspace{-0.3cm}
  \includegraphics[width=5.5cm]{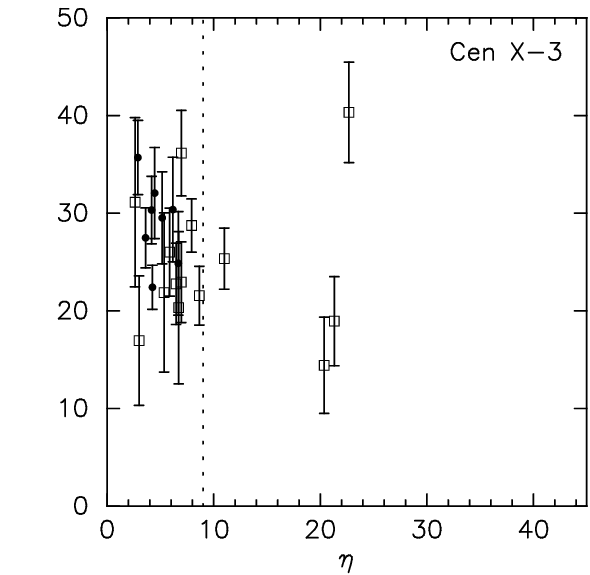}
  \caption{The upper panels show the radial-velocity amplitude
  ($K_{\rm opt}$ in \kms) measured for a given line as a function of
  its mean line equivalent width ($<{\rm EW}>$). The error bars
  indicate 1 $\sigma$ errors. Especially for \object{SMC~X$-$1} a
  clear trend is visible with line strength: for large mean EW $K_{\rm
  opt}$ is systematically lower. The finally adopted $K_{\rm opt}$ is
  represented by the horizontal dotted line. The middle panels show
  the $\gamma$-velocity and the bottom panels show $K_{\rm opt}$ as a
  function of the line variability parameter $\eta$; lines with $\eta
  > 9$ are rejected (see text for motivation). The lines that are rejected 
  because of this and
  other criteria (such as blending) are represented by open
  squares. The filled circles indicate the lines that are used to
  determine the (mean) radial-velocity amplitude. These selections
  show less spread in $K_{\rm opt}$.}
  \label{ew_plots4}
  \end{center} 
  \end{figure*}

  Cosmic ray hits were removed by rejecting the affected wavelength
  bins and subsequent interpolation. After this, the spectra were
  normalised by fitting the continuum with a spline over the whole
  wavelength range of one spectral arm. The normalised spectra of the
  three systems are shown in Figs.~\ref{spectra_blue4} to
  \ref{spectra_red24}.

  We verified the long-term stability of UVES by measuring the position
  of the interstellar lines of $\ion{Ca}{ii}$~K at 3933.66~{\AA},
  $\ion{Ca}{ii}$~H at 3968.47~{\AA}, $\ion{Na}{i}$~D1 at 5895.92~{\AA}
  and $\ion{Na}{i}$~D2 at 5889.95~{\AA} for all spectra. This resulted
  in a deviation of less than 1~\kms\ throughout the whole
  observing period of each system, i.e. several months.
  \begin{table*}[!t]
  \caption[]{List of lines that have been fitted with a gaussian. The
  rest wavelength of the line is given as well as the mean equivalent
  width (EW, in \AA ngstr\"{o}m) with its $1\sigma$ error. The
  magnitude of the variations in line EW is expressed as the ratio
  $\eta$ of the standard deviation of the line EW to the error on the
  mean EW. Rejected lines are indicated with a remark: lines marked
  with an ``B'' are blends; a ``V'' stands for lines exhibiting strong
  EW variations ($\eta > 9$); a number (e.g. 0.5) indicates a
  deviation in the EW at that orbital phase. Electronic tables that
  contain the radial-velocity amplitude, EW and FWHM variations for
  each line are available as on-line material at the CDS.}
  \label{linelist4}
  \begin{center}
  \begin{tabular}{ll | lll | lll | lll }
  \hline \hline
  & & \multicolumn{3}{|c|}{\object{SMC~X$-$1}}
  & \multicolumn{3}{|c|}{\object{LMC~X$-$4}}
  & \multicolumn{3}{|c}{\object{Cen~X$-$3}}
  \\
  \hline
  Identification & $\lambda_{\rm rest}$ ({\AA})
                 & $\mu_{\rm EW}$ ({\AA}) & $\eta$ & Remarks
                 & $\mu_{\rm EW}$ ({\AA}) & $\eta$ & Remarks
                 & $\mu_{\rm EW}$ ({\AA}) & $\eta$ & Remarks \\
  \hline
  \multicolumn{2}{l|}{\textit{H Balmer series}} & & & & & & & & & \\
  \ion{H$\beta$}{}    & $ 4861.33 $ & $ 1.265(4)  $ & $ 44.1 $ & V,0.5    &   $ 1.905(7)   $ & $ 22.6 $ & V,0.5    & $ 1.491(6)   $ & $ 20.4 $ & V,0.5     \\
  \ion{H$\gamma$}{}   & $ 4340.46 $ & $ 1.287(5)  $ & $ 24.2 $ & B,V,0.5  &   $ 1.796(8)   $ & $ 14.9 $ & B,V,0.5  & $ 1.638(10)  $ & $  8.6 $ & B,0.5     \\
  \ion{H$\delta$}{}   & $ 4101.73 $ & $ 1.503(7)  $ & $ 17.5 $ & B,V,0.5  &   $ 2.216(9)   $ & $  7.6 $ & B,V,0.5  & $ 2.264(16)  $ & $  7.0 $ & B,0.5     \\
  \ion{H$\epsilon$}{} & $ 3970.07 $ & $ 1.640(13) $ & $  8.5 $ & B,0.5    &   $ 1.742(11)  $ & $ 10.4 $ & B,V,0.5  & $ 1.356(23)  $ & $ 22.7 $ & B,V,0.5   \\
  \ion{H8}{}          & $ 3889.05 $ & $ 1.173(4)  $ & $  9.7 $ & V,0.5    &   $ 1.538(6)   $ & $ 10.0 $ & V,0.5    & $ 1.278(11)  $ & $  6.5 $ & 0.5       \\
  \ion{H9}{}          & $ 3835.38 $ & $ 1.260(5)  $ & $  9.2 $ & B?,V,0.5 &   $ 1.523(7)   $ & $  9.2 $ & B?,V     & $ 1.211(12)  $ & $  7.9 $ & B?,0.5    \\
  \ion{H10}{}         & $ 3797.90 $ & $ 1.039(5)  $ & $  8.2 $ &          &   $ 1.259(7)   $ & $  8.4 $ &          & $ 0.843(12)  $ & $  5.9 $ & 0.5       \\
  \ion{H11}{}         & $ 3770.63 $ & $ 1.007(6)  $ & $  8.6 $ &          &   $ 1.200(10)  $ & $  6.8 $ &          & $ 0.980(18)  $ & $  7.0 $ & 0.5       \\
  \ion{H12}{}         & $ 3750.15 $ & $ 0.779(6)  $ & $  6.6 $ &          &   $ 0.638(10)  $ & $  7.4 $ & B        & $ 0.413(15)  $ & $  5.3 $ & B         \\
  \ion{H13}{}         & $ 3734.37 $ & $ 0.745(6)  $ & $  6.5 $ &          &   $ 0.588(12)  $ & $  6.1 $ &          & $ 0.548(22)  $ & $  4.5 $ &           \\
  \ion{H14}{}         & $ 3721.94 $ & $ 0.513(8)  $ & $  5.9 $ & B        &   $ -          $ & $  -   $ &          & $ -          $ & $  -   $ &           \\
  \ion{H15}{}         & $ 3711.97 $ & $ 0.470(11) $ & $  4.6 $ &          &   $ -          $ & $  -   $ &          & $ -          $ & $  -   $ &           \\
  \ion{H16}{}         & $ 3703.85 $ & $ 0.545(8)  $ & $  8.3 $ & B        &   $ 0.764(237) $ & $  3.1 $ & B        & $ 0.466(96)  $ & $  3.0 $ & B         \\[0.15cm]
 		     		      	               			   	   	    	 	       	     	      	    	      	 	       	      			  			       	   		      			   		     	       	            	  				  	      							 		     		       
  \multicolumn{2}{l|}{\textit{He I 3P-3D series}} & & & & & & & & & \\	   	   	    	 	       	     	      	    	      	 	       	      
  \ion{He}{i} 2p-3d   & $ 5875.66 $ & $ 0.981(5)  $ & $ 31.3 $ & V,0.0    &   $ 1.159(11)  $ & $  8.6 $ & 0.0      & $ 1.194(7)   $ & $ 21.3 $ & V,0.0,0.5 \\
  \ion{He}{i} 2p-4d   & $ 4471.50 $ & $ 0.773(6)  $ & $ 12.3 $ & V,0.0    &   $ 0.775(8)   $ & $  8.1 $ &          & $ 0.739(10)  $ & $  6.7 $ &           \\
  \ion{He}{i} 2p-5d   & $ 4026.21 $ & $ 0.697(4)  $ & $ 12.1 $ & V,0.0    &   $ 0.825(5)   $ & $  8.7 $ &          & $ 0.653(8)   $ & $  3.6 $ &           \\
  \ion{He}{i} 2p-6d   & $ 3819.62 $ & $ 0.509(4)  $ & $ 11.3 $ & V,0.0    &   $ 0.460(5)   $ & $  8.9 $ &          & $ 0.313(10)  $ & $  2.9 $ &           \\
  \ion{He}{i} 2p-7d   & $ 3705.02 $ & $ 0.476(9)  $ & $  8.2 $ & B,0.0    &   $ 0.632(66)  $ & $  4.4 $ & B        & $ 0.556(152) $ & $  2.6 $ & B         \\
  \ion{He}{i} 2p-8d   & $ 3634.25 $ & $ 0.244(5)  $ & $  5.7 $ &          &   $ 0.208(7)   $ & $  6.4 $ &          & $ -          $ & $  -   $ &           \\
  \ion{He}{i} 2p-9d   & $ 3587.27 $ & $ 0.169(5)  $ & $  7.2 $ &          &   $ -          $ & $  -   $ &          & $ -          $ & $  -   $ &           \\[0.15cm]
		     		      	               			   	   	    	 	       	     	      	    	      	 	       	      			  			       	   		      			   		     	       	            	  				  	      							 		     		       
  \multicolumn{2}{l|}{\textit{He I 1P-1D series}} & & & & & & & & & \\	   	   	    	 	       	     	      	    	      	 	       	      
  \ion{He}{i} 2p-4d   & $ 4921.93 $ & $ 0.595(3)  $ & $ 16.2 $ & V,0.0    &   $ 0.530(5)   $ & $  7.0 $ &          & $ 0.323(5)   $ & $  5.2 $ &           \\
  \ion{He}{i} 2p-5d   & $ 4387.93 $ & $ 0.398(5)  $ & $  6.5 $ &          &   $ 0.302(6)   $ & $  5.5 $ &          & $ -          $ & $  -   $ &           \\
  \ion{He}{i} 2p-6d   & $ 4143.76 $ & $ 0.341(4)  $ & $  6.1 $ &          &   $ 0.301(6)   $ & $  6.4 $ &          & $ -          $ & $  -   $ &           \\
  \ion{He}{i} 2p-7d   & $ 4009.26 $ & $ 0.238(4)  $ & $  4.8 $ &          &   $ -          $ & $  -   $ &          & $ -          $ & $  -   $ &           \\[0.15cm]
				      	               			   	   	    	 	       	     	      	    	      	 	       	      			  			       	   		      			   		     	       	            	  				  	      							 		     		       
  \multicolumn{2}{l|}{\textit{He I 3P-3S series}} & & & & & & & & & \\	   	   	    	 	       	     	      	    	      	 	       	      
  \ion{He}{i} 2p-4s   & $ 4713.17 $ & $ 0.311(4)  $ & $  6.6 $ &          &   $ 0.265(5)   $ & $  4.6 $ &          & $ 0.228(6)   $ & $  6.2 $ &           \\[0.15cm]
				      	               			   	   	    	 	       	     	      	    	      	 	       	      			  			       	   		      			   		     	       	            	  				  	      							 		     		       
  \multicolumn{2}{l|}{\textit{He I 1S-1P series}} & & & & & & & & & \\	   	   	    	 	       	     	      	    	      	 	       	      
  \ion{He}{i} 2s-3p   & $ 5015.68 $ & $ 0.353(3)  $ & $ 10.9 $ & V,0.0    &   $ 0.307(5)   $ & $  7.4 $ &          & $ 0.310(5)   $ & $  4.2 $ &           \\[0.15cm]
				      	               			   	   	    	 	       	     	      	    	      	 	       	      			  			       	   		      			   		     	       	            	  				  	      							 		     		       
  \multicolumn{2}{l|}{\textit{He I 1P-1S series}} & & & & & & & & & \\	   	   	    	 	       	     	      	    	      	 	       	      
  \ion{He}{i} 2p-4s   & $ 5047.74 $ & $ 0.201(4)  $ & $  6.7 $ &          &   $ -          $ & $  -   $ &          & $ -          $ & $  -   $ &           \\[0.15cm]
				      	               			   	   	    	 	       	     	      	    	      	 	       	      			  			       	   		      			   		     	       	            	  				  	      							 		     		       
  \multicolumn{2}{l|}{\textit{Other lines}} & & & & & & & \\			   	   	    	 	       	     	      	    	      	 	       	      
  \ion{He}{ii} 4-7    & $ 5411.53 $ & $ 0.292(4)  $ & $ 42.5 $ & V,0.5    &   $ 0.809(7)   $ & $ 14.0 $ & V,0.5    & $ 0.827(5)   $ & $ 11.0 $ & B,0.5     \\
  \ion{He}{ii} 4-11   & $ 4199.83 $ & $ 0.145(5)  $ & $  9.7 $ & V,0.5    &   $ 0.485(7)   $ & $  5.6 $ &          & $ 0.494(9)   $ & $  4.2 $ &           \\
  \ion{Si}{iv} 4s-4p  & $ 4088.86 $ & $ 0.310(4)  $ & $ 10.3 $ & V,0.0    &   $ 0.300(11)  $ & $  6.6 $ & B        & $ 0.326(12)  $ & $  6.7 $ & B         \\
  \hline
  \end{tabular}
  \end{center}
  \end{table*}


\section{Spectral Analysis}
\label{spectra4}

  To obtain a radial-velocity measurement often the complete spectrum
  is cross-correlated with a template spectrum. This approach has
  many advantages when using spectra with relatively low spectral
  resolution and poor signal-to-noise. In our case the spectra are of
  such high quality that the radial-velocity amplitude can be
  determined for each line separately. The advantage of such a
  strategy is that it is possible to assess the influence of possible
  distortions due to e.g. X-ray heating and gravity darkening, as in
  these systems the OB star is irradiated by a powerful X-ray source
  ($L_{\rm opt} \simeq L_{X}$) and is filling its
  Roche-lobe. Furthermore, the extended OB-star wind is focused into
  a shadow wind which possibly produces a strong shock (a so-called
  photo-ionisation wake, see e.g. \citealt{blo94,kap94}) where the
  fast shadow wind catches up with the stagnant flow inside the X-ray
  ionization zone. The shadow wind and photo-ionization wake induce
  orbital modulations of spectral lines formed in the stellar wind
  (i.e. strong spectral lines such as the first lines of the Balmer
  series and the strongest helium lines).

  Figures~\ref{spectra_blue4} to \ref{spectra_red24} show that the
  spectra contain mostly lines that are identified with transitions
  from \ion{H}{} and \ion{He}{i}. Only a few \ion{He}{ii} lines and
  some metal lines are detected, consistent with the modest
  metallicity of the Magellanic Clouds and OB supergiant spectral
  types. We show the spectra observed near X-ray eclipse (orbital
  phase $\phi \sim 0.0$). Some of the lines are blended or show a
  slight asymmetry. A comparison of the spectra obtained at different
  orbital phase reveals that several lines also vary in line
  strength. Still, Fig.~\ref{gaussfits4} demonstrates that many lines
  are well represented by a gaussian profile (i.e. as one would expect
  for a rotationally broadened profile). Therefore, gaussian profiles
  are fit to all individual lines to determine the radial-velocity
  curve (Sect.~\ref{radvel_curves4}). Subsequently, the observed line
  profiles are examined on asymmetry and variations with orbital phase
  (Sect.~\ref{line_selection4}).
  \begin{figure}[!t]
  \begin{center}
  \includegraphics[width=7.7cm]{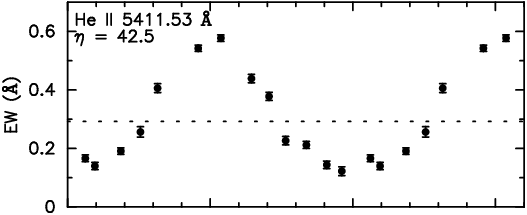}
  \includegraphics[width=7.7cm]{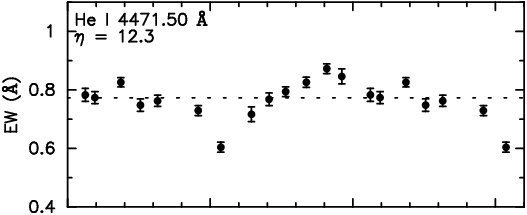}
  \includegraphics[width=7.7cm]{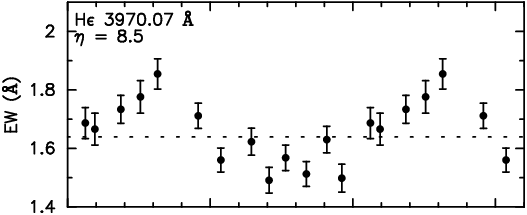}
  \includegraphics[width=7.7cm]{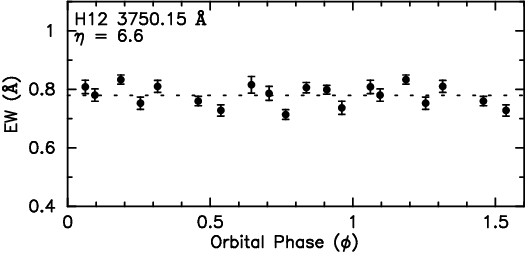}
  \caption{Equivalent width (EW) variations with orbital phase
  ($\phi$) observed in \object{SMC~X$-$1}. The error bars indicate 1
  $\sigma$ errors and the dotted line the average value. Some
  datapoints are shown twice to better visualise trends with orbital
  phase. The value of the defined line variability parameter $\eta$ is
  indicated for each line. The upper panel shows the large variations
  in the \ion{He}{ii} line at 5411.53~{\AA}, the second panel shows
  the inverse behaviour of the \ion{He}{i} line at 4471.50~{\AA}, the
  third panel shows the variations of \ion{H$\epsilon$}{} at
  3970.07~{\AA}. A stable line is shown in the lower panel, i.e. the
  \ion{H12}{} line at 3750.15~{\AA}. The EW ratio of the \ion{He}{ii}
  and \ion{He}{i} lines is sensitive to $T_{\rm eff}$ and can be used
  for spectral classification; apparently, the spectral type of
  \object{SMC~X$-$1} varies with orbital phase. The variations are
  consistent with being caused by X-ray heating of the stellar surface
  facing the X-ray source.}
  \label{ew_var4}
  \end{center}
  \end{figure}

\subsection{Radial-velocity curves}
\label{radvel_curves4}

  We determine the line centre, and thus the Doppler shift with
  respect to the heliocentric restframe, by fitting the profile with a
  gaussian. The gaussian sets the full-width at half maximum (FWHM),
  the central line depth and the central wavelength of the profile,
  i.e. three free parameters. A $\chi^{2}$ minimalization procedure
  delivers the best fit gaussian profile and defines the accuracy of
  the fit parameters. Emission-line profiles, such as H$\alpha$, are
  not included in the fitting procedure.

  The orbital parameters are accurately known from X-ray pulse time
  delay measurements (see Table~\ref{literature_values4}). We assume
  the orbit to be circular, since in all cases the eccentricity $e
  \lesssim 0.008$. The remaining free parameters describing the
  radial-velocity curve are the radial-velocity amplitude, $K_{\rm
  opt}$, and the restframe velocity of the system, $v_{\gamma}$. In
  principle, a small shift in orbital phase could be present due to
  the inaccuracy of the orbital period, the period derivative, and
  mid-eclipse time (Table~\ref{literature_values4}).  The measured
  orbital phase shifts are not significantly different from zero, but
  are slightly larger than the phase shifts one may expect based on
  the accuracy of the ephemeres of these systems. The phase shift
  should be the same for all lines and has to be fixed to the average
  value during a second iteration.

  A radial-velocity curve is obtained for each individual line;
  Fig.~\ref{radvels_oneline4} displays for each system a
  radial-velocity curve representative for the spectral lines
  used to measure the radial-velocity amplitude. Note that
  the data points were obtained from several orbits of the system; the
  dozen spectra per system are evenly distributed with orbital phase,
  thanks to the service-mode observations allowing to obtain spectra
  spread over a period of more than one month.

  Fig.~\ref{ew_plots4} indicates that $K_{\rm opt}$ and $v_{\gamma}$
  show quite some dispersion when comparing the radial-velocity curves
  of individual lines. $K_{\rm opt}$ shows a dependence on line
  strength and line variability, diagnostics that we will use to
  reject lines when determining the mean radial-velocity amplitude
  used to calculate the mass of the neutron star.
  \begin{figure}[!t]
  \begin{center}
  \includegraphics[width=7.7cm]{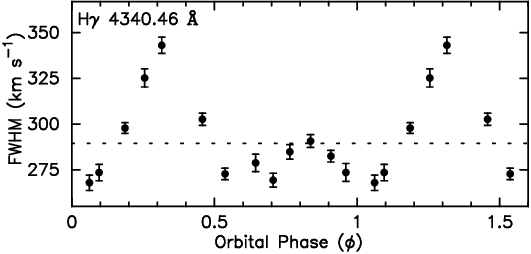}

  \vspace{0.1cm}
  \includegraphics[width=7.7cm]{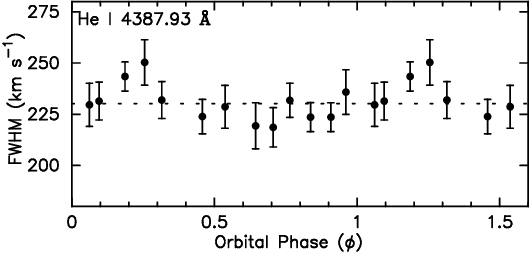}
  \caption{Full width at half maximum (FWHM) variations of the
  \ion{H$\gamma$}{} line at 4340.46~{\AA} (upper panel) and the
  \ion{He}{i} line at 4387.93~{AA} (lower panel) of \object{SMC~X$-$1}
  as a function of orbital phase ($\phi$) in \kms. The error bars
  indicate 1 $\sigma$ errors and the dotted line the average
  value. Some datapoints are repeated once to better visualise trends
  with orbital phase. A maximum in FWHM is clearly seen around $\phi
  \sim 0.25$; a second, though much less pronounced maximum is visible
  at $\phi \sim 0.75$. Such FWHM variations are also detected in other
  lines, as well as in the systems \object{LMC~X$-$4} and
  \object{Cen~X$-$3}.}
  \label{fwhm_var4}
  \end{center}
  \end{figure}

\subsection{Line Selection}
\label{line_selection4}

  As our observational strategy is aimed at the derivation of the
  radial-velocity curve based on individual lines, we select
  only lines that are well identified. Also, all lines that are
  (partially) blended, are rejected. In Table~\ref{linelist4} this is
  indicated with a ``B'' as a remark.

  In these high-mass X-ray binaries it is expected that at least some
  lines in the OB-star spectrum are affected by the presence of the
  X-ray pulsar companion (see e.g. \citealt{par78,rey93}). In
  principle, one could model the photospheric line profiles of the OB
  supergiant companion by calculating a grid of many (thousands)
  surface elements, adopting for each surface element an intrinsic
  line profile, and obtaining the integrated line profile by a
  weighted integration of the (mainly by stellar rotation)
  Doppler-shifted intrinsic profiles of all visible surface
  elements. Such computationally demanding techniques have been
  successfully explored in stellar pulsation studies
  (e.g. \citealt{sch97}) and in modeling photometric lightcurves of
  HMXBs (e.g. \citealt{hee89}). Also, variations in the intrinsic line
  profiles due to variations of surface temperature and gravity have
  been included \citep{sch99}, so that one could apply such a method
  to the case of an OB supergiant irradiated and deformed by a close
  and compact X-ray source. Although we have now made an advance in
  analysing spectra of the OB companions to X-ray pulsars by studying
  individual lines rather than cross-correlating complete spectra, we
  consider this modeling effort beyond the scope of the present paper,
  even though it has the potential to deliver very important
  information on the interpretation and analysis of the obtained
  spectra.
  \begin{table*}[!ht]
  \caption[]{Final selection of lines that is used for the
  determination of the radial-velocity amplitude ($K_{\rm opt}$) for each
  system. The three fit parameters are listed, i.e. the system
  velocity ($v_\gamma$), $K_{\rm opt}$, and a phase shift ($\Delta
  \phi$). The goodness of the fit is expressed as $\chi^2_r$. All
  errors are 1 $\sigma$.}
  \label{fitparameters4}
  \begin{center}
  \begin{footnotesize}
  \begin{tabular}{ p{1.0cm} | p{1.4cm} p{1.2cm} p{0.9cm} p{0.3cm} | p{1.4cm} p{1.2cm} p{0.9cm} p{0.3cm} | p{1.4cm} p{1.2cm} p{0.9cm} p{0.3cm} }
  \hline \hline
  & \multicolumn{4}{|c|}{\object{SMC~X$-$1}}
  & \multicolumn{4}{|c|}{\object{LMC~X$-$4}}
  & \multicolumn{4}{|c}{\object{Cen~X$-$3}}
  \\
  \hline
  $\lambda_{\rm rest}$ ({\AA}) &
  $v_\gamma$ & $K_{\rm opt}$ & $\Delta \phi$ & $\chi^2_r$ &
  $v_\gamma$ & $K_{\rm opt}$ & $\Delta \phi$ & $\chi^2_r$ &
  $v_\gamma$ & $K_{\rm opt}$ & $\Delta \phi$ & $\chi^2_r$ \\
  \hline
  $ 3797.90 $ & $ 190.0 \pm 0.9 $ & $ 20.2 \pm 1.2 $ & $ 0.01(1) $ & $ 2.6 $ & $ 305.5 \pm 1.1  $ & $ 27.6 \pm 1.6  $ & $  0.01(1) $ & $ 1.6 $ & $ -            $ & $ -            $ & $ -       $ & $ -   $ \\
  $ 3770.63 $ & $ 191.7 \pm 1.3 $ & $ 18.1 \pm 1.8 $ & $ 0.01(2) $ & $ 3.8 $ & $ 321.5 \pm 2.1  $ & $ 32.6 \pm 2.9  $ & $ -0.02(2) $ & $ 3.2 $ & $ -            $ & $ -            $ & $ -       $ & $ -   $ \\
  $ 3750.15 $ & $ 184.5 \pm 1.6 $ & $ 18.4 \pm 2.2 $ & $ 0.04(2) $ & $ 4.3 $ & $ -              $ & $ -             $ & $  -       $ & $ -   $ & $ -            $ & $ -            $ & $ -       $ & $ -   $ \\
  $ 3734.37 $ & $ 186.5 \pm 1.8 $ & $ 18.1 \pm 2.5 $ & $ 0.02(2) $ & $ 4.9 $ & $ 306.2 \pm 2.6  $ & $ 33.6 \pm 3.8  $ & $  0.02(2) $ & $ 1.7 $ & $ 39.5 \pm 4.1 $ & $ 32.1 \pm 4.7 $ & $ 0.02(4) $ & $ 0.9 $ \\
  $ 3711.97 $ & $ 207.9 \pm 1.8 $ & $ 20.4 \pm 2.4 $ & $ 0.07(2) $ & $ 0.9 $ & $ -              $ & $ -             $ & $  -       $ & $ -   $ & $ -            $ & $ -            $ & $ -       $ & $ -   $ \\
  $ 4471.50 $ & $ -             $ & $ -            $ & $ -       $ & $ -   $ & $ 298.9 \pm 1.9  $ & $ 40.1 \pm 2.7  $ & $  0.00(1) $ & $ 2.7 $ & $ 20.5 \pm 3.9 $ & $ 24.9 \pm 5.3 $ & $ 0.08(4) $ & $ 5.8 $ \\
  $ 4026.21 $ & $ -             $ & $ -            $ & $ -       $ & $ -   $ & $ 290.2 \pm 1.3  $ & $ 35.5 \pm 1.8  $ & $  0.00(1) $ & $ 2.9 $ & $  9.9 \pm 2.6 $ & $ 27.5 \pm 3.1 $ & $ 0.03(3) $ & $ 3.3 $ \\
  $ 3819.62 $ & $ -             $ & $ -            $ & $ -       $ & $ -   $ & $ 301.9 \pm 1.7  $ & $ 40.0 \pm 2.5  $ & $ -0.01(1) $ & $ 2.0 $ & $ 38.8 \pm 2.8 $ & $ 35.7 \pm 3.8 $ & $ 0.07(2) $ & $ 0.7 $ \\
  $ 3634.25 $ & $ 187.7 \pm 2.4 $ & $ 20.5 \pm 3.2 $ & $ 0.03(3) $ & $ 2.0 $ & $ 312.1 \pm 2.9  $ & $ 50.0 \pm 4.3  $ & $ -0.03(2) $ & $ 0.8 $ & $ -            $ & $ -            $ & $ -       $ & $ -   $ \\
  $ 3587.27 $ & $ 183.9 \pm 2.1 $ & $ 19.0 \pm 2.8 $ & $ 0.01(3) $ & $ 1.0 $ & $ -              $ & $ -             $ & $  -       $ & $ -   $ & $ -            $ & $ -            $ & $ -       $ & $ -   $ \\
  $ 4921.93 $ & $ -             $ & $ -            $ & $ -       $ & $ -   $ & $ 311.7 \pm 1.5  $ & $ 34.8 \pm 2.1  $ & $ -0.02(1) $ & $ 1.8 $ & $ 44.3 \pm 3.3 $ & $ 29.5 \pm 4.7 $ & $ 0.10(3) $ & $ 3.0 $ \\
  $ 4387.93 $ & $ 193.9 \pm 1.8 $ & $ 23.7 \pm 2.4 $ & $ 0.04(2) $ & $ 3.1 $ & $ 328.6 \pm 2.6  $ & $ 38.9 \pm 3.6  $ & $ -0.01(2) $ & $ 1.4 $ & $ -            $ & $ -            $ & $ -       $ & $ -   $ \\
  $ 4143.76 $ & $ 194.2 \pm 2.1 $ & $ 25.6 \pm 2.7 $ & $ 0.02(2) $ & $ 3.7 $ & $ 315.3 \pm 3.1  $ & $ 50.9 \pm 4.4  $ & $ -0.01(2) $ & $ 1.6 $ & $ -            $ & $ -            $ & $ -       $ & $ -   $ \\
  $ 4009.26 $ & $ 190.1 \pm 1.9 $ & $ 20.4 \pm 2.5 $ & $ 0.05(2) $ & $ 1.8 $ & $ -              $ & $ -             $ & $  -       $ & $ -   $ & $ -            $ & $ -            $ & $ -       $ & $ -   $ \\
  $ 4713.17 $ & $ 190.9 \pm 1.3 $ & $ 22.3 \pm 1.9 $ & $ 0.02(1) $ & $ 2.3 $ & $ 315.7 \pm 2.7  $ & $ 39.8 \pm 4.0  $ & $  0.00(2) $ & $ 1.7 $ & $ 24.6 \pm 4.0 $ & $ 30.4 \pm 5.4 $ & $ 0.08(3) $ & $ 1.6 $ \\
  $ 5047.74 $ & $ 183.5 \pm 2.8 $ & $ 21.7 \pm 3.9 $ & $ 0.03(3) $ & $ 3.0 $ & $ -              $ & $ -             $ & $  -       $ & $ -   $ & $ -            $ & $ -            $ & $ -       $ & $ -   $ \\
  $ 5015.68 $ & $ -             $ & $ -            $ & $ -       $ & $ -   $ & $ 312.7 \pm 2.4  $ & $ 39.5 \pm 3.2  $ & $ -0.03(2) $ & $ 2.1 $ & $ 26.9 \pm 2.5 $ & $ 30.3 \pm 3.5 $ & $ 0.09(2) $ & $ 1.7 $ \\
  $ 4199.83 $ & $ -             $ & $ -            $ & $ -       $ & $ -   $ & $ 303.3 \pm 1.3  $ & $ 26.6 \pm 2.0  $ & $  0.05(1) $ & $ 0.6 $ & $ 44.4 \pm 2.0 $ & $ 22.4 \pm 2.3 $ & $ 0.03(2) $ & $ 0.8 $ \\
  \hline
  \end{tabular}
  \end{footnotesize}
  \end{center}
  \end{table*}
  
  A first step in this direction has been undertaken by \citet{abu04}; 
  they model the line formation process taking X-ray heating
  and gravitational darkening into account. As the luminosity
  distribution and shape of the star are altered, the measured radial
  velocity, as well as the line equivalent width, depend on the depth
  of the line-forming region and orbital phase. \citet{abu04} show
  that these effects will result in a reduction of the derived
  radial-velocity amplitude by several \kms.

  That these effects manifest themselves in our observations is nicely
  demonstrated by the large EW variations of the \ion{He}{ii} line at
  5411.53~{\AA} of \object{SMC~X$-$1} (see Fig.~\ref{ew_var4}). This
  line varies with about a factor $\sim 5$ in EW and reaches a maximum
  EW near $\phi \sim 0.5$. The \ion{He}{i} line at 4471.50~{\AA} (and
  also \ion{He}{i} 5875.66~{\AA}) shows the opposite behaviour and has
  a minimum strength at this orbital phase (see
  Fig.~\ref{ew_var4}). Since a \ion{He}{i} / \ion{He}{ii} line
  ratio is sensitive to $T_{\rm eff}$ it can be used for spectral
  classification; however, traditionally the ratio of \ion{He}{i} 4471
  over \ion{He}{ii} 4541, accessible in the blue spectrum, is used for
  spectral classification \citep{con71,len93,mok05}.  Line ratios
  involving the \ion{He}{i} 5876 and \ion{He}{ii} 5412 lines have not
  (yet) been calibrated (Mokiem, priv. comm.). The observed
  variations in line ratio indicate a higher $T_{\rm eff}$ at the side
  of the OB star facing the X-ray source, and is evidence for X-ray
  heating in \object{SMC~X$-$1}. As a consequence, the spectral type
  of the optical companion to \object{SMC~X$-$1} changes as function
  of binary aspect angle. A similar effect has been observed in the
  optical spectrum of \object{V779~Cen}, the O6 companion of
  \object{Cen~X$-$3} \citep{hut79}.

  The derived value of the radial-velocity amplitude shows a
  dependence on line strength (Fig.~\ref{ew_plots4}). Furthermore, the
  observed \textit{variations} in EW also increase with line strength. As
  illustrated above, these variations likely reflect distortions of
  the line forming region due to e.g. X-ray heating and the
  (disturbed) stellar wind. Since $K_{\rm opt}$ should be a unique
  value, we investigate whether a selection criterion can be defined
  to reject lines that are affected by these distortions and thus do
  not yield a sound measurement of $K_{\rm opt}$.
  \begin{table*}[!t]
  \caption[]{Weighted mean values of the radial velocity for each
  spectrum, indicated by its orbital phase ($\phi$) for
  \object{SMC~X$-$1}, \object{LMC~X$-$4}, and \object{Cen~X$-$3}. All
  errors are 1 $\sigma$.}
  \label{mean_velocities4}
  \begin{center}
  \begin{tabular}{ p{1.0cm} p{2.4cm}  | p{1.0cm} p{2.4cm} | p{1.0cm} p{2.4cm} }
  \hline \hline
  \multicolumn{2}{c|}{\object{SMC~X$-$1}}
  & \multicolumn{2}{|c|}{\object{LMC~X$-$4}}
  & \multicolumn{2}{|c}{\object{Cen~X$-$3}}
  \\
  \hline
  $\phi$ & $<v_{\rm rad}>$ & $\phi$ & $<v_{\rm rad}>$ & $\phi$ & $<v_{\rm rad}>$ \\
         & (\kms)            &        & (\kms)          &         & (\kms) \\
  \hline
  $ 0.062 $ & $   8.42 \pm 1.06 $ & $ 0.037 $ & $   6.45 \pm 1.50 $ & $ 0.095 $ & $  14.96 \pm 2.97 $ \\
  $ 0.096 $ & $  19.04 \pm 1.02 $ & $ 0.138 $ & $  29.28 \pm 2.23 $ & $ 0.234 $ & $  22.87 \pm 2.10 $ \\
  $ 0.186 $ & $  16.58 \pm 0.77 $ & $ 0.247 $ & $  40.06 \pm 1.38 $ & $ 0.284 $ & $  31.47 \pm 1.98 $ \\
  $ 0.255 $ & $  24.42 \pm 1.17 $ & $ 0.274 $ & $  30.03 \pm 1.60 $ & $ 0.337 $ & $  14.12 \pm 2.59 $ \\
  $ 0.315 $ & $  14.70 \pm 0.95 $ & $ 0.337 $ & $  24.84 \pm 1.24 $ & $ 0.354 $ & $  18.43 \pm 2.75 $ \\
  $ 0.458 $ & $   4.98 \pm 0.84 $ & $ 0.388 $ & $  25.04 \pm 1.33 $ & $ 0.462 $ & $ -10.75 \pm 2.86 $ \\
  $ 0.537 $ & $  -9.37 \pm 1.01 $ & $ 0.418 $ & $  23.23 \pm 1.10 $ & $ 0.512 $ & $ -21.61 \pm 2.42 $ \\
  $ 0.645 $ & $ -18.46 \pm 1.24 $ & $ 0.533 $ & $ -14.69 \pm 1.54 $ & $ 0.681 $ & $ -26.12 \pm 2.19 $ \\
  $ 0.707 $ & $ -19.07 \pm 1.08 $ & $ 0.694 $ & $ -30.90 \pm 1.21 $ & $ 0.746 $ & $ -18.81 \pm 1.76 $ \\
  $ 0.765 $ & $ -16.47 \pm 0.89 $ & $ 0.743 $ & $ -34.06 \pm 1.17 $ & $ 0.761 $ & $ -23.29 \pm 2.07 $ \\
  $ 0.837 $ & $ -19.83 \pm 0.87 $ & $ 0.809 $ & $ -30.86 \pm 1.16 $ & $ 0.831 $ & $ -24.32 \pm 2.34 $ \\
  $ 0.909 $ & $  -8.22 \pm 0.77 $ & $ 0.927 $ & $ -17.96 \pm 1.19 $ & $ 0.515 $ & $ -14.55 \pm 1.98 $ \\
  $ 0.961 $ & $  -1.85 \pm 1.09 $ & $ 0.927 $ & $ -17.95 \pm 1.12 $ & $       $ & $                 $ \\
  \hline
  \end{tabular}
  \end{center}
  \end{table*}

  To measure these distortions we apply a velocity moment analysis,
  with which one can determine the equivalent width (EW), central
  velocity, standard deviation ($\sigma$) and skewness ($\tau$) of a
  given line. The n$^{\rm th}$ moment ($\mu_n$) of a distribution
  $f(v)$ in velocity $v$ is given by:
  \begin{equation}
    \mu_n = \frac{\int{v^n~f(v)~dv}}{\int{f(v)~dv}},
  \end{equation}
  and
  \begin{equation} 
    \mu_0 = \int{f(v)~dv}.
  \end{equation}
  For such a distribution the EW is proportional to
  $\mu_0$ and the central velocity to $\mu_1$. The standard deviation
  ($\sigma$) and skewness ($\tau$) are often defined as:
  \begin{equation} 
    \sigma^2 = \frac{\int{(v - \mu_1)^2~f(v)~dv}}{\mu_0},
  \end{equation}
  and
  \begin{equation} 
    \tau = \frac{\int{(v - \mu_1)^3~f(v)~dv}}{\mu_0~\sigma^3},
    \label{equation_tau4}
  \end{equation}
  in which the skewness ($\tau$) is a measure of the asymmetry of the
  distribution. For a gaussian profile the skewness $\tau $ is zero;
  if $\tau \neq 0$ the spectral line is not well represented by a
  gaussian. If so, one can not use a gaussian to fit the line in order
  to measure its radial velocity, as we did in
  Sect.~\ref{radvel_curves4}. It turns out that for the lines
  that are not blended, $\tau$ is consistent with being equal to zero
  within the error. Note, however, that the error on $\tau$ is too
  large to measure any significant deviations from zero, because it
  depends on the error on $\mu$ and $\sigma$ to the third power (see
  Eq.~\ref{equation_tau4}). The EW and central velocity (first moment)
  are much more accurately determined and are consistent with the
  respective values derived from the gaussian profile fits.

  For each line we determine whether the line EW varies significantly
  by comparing the deviation in EW to the error on the mean EW
  (Table~\ref{linelist4}). Note that these values are not equal; the
  error on the mean depends on the error in EW of the individual
  spectra, while the standard deviation is the spread in the
  \textit{distribution} of EWs. Since the strongest lines are formed
  in the outer layers of the stellar photosphere and/or in the
  extended stellar wind these lines are expected to be most affected
  by X-ray heating, gravity darkening, etc., and will thus show
  intrinsic variations when the system revolves. The observed trend in
  $K_{\rm opt}$ with $v_{\gamma}$ (Fig.~\ref{ew_plots4}) also
  indicates that the stronger lines are formed further out in the
  stellar atmosphere and wind (an effect called Balmer progression,
  see e.g. \citealt{cra85,abu04}). We define a line variability
  parameter $\eta$ to formulate a selection criterion
  (Table~\ref{linelist4}). This parameter is defined as the ratio of
  the standard deviation of the EW variations to the error on the mean
  EW. The main motivation behind this definition is that the EW is in
  principle not sensitive to radial-velocity variations, i.e. the key
  parameter that we want to measure in the spectra. However, $\eta$
  turns out to be an accurate probe of intrinsic line profile
  variability. Other methods to measure line profile variability
  (e.g. the temporal variance spectrum analysis method introduced by
  \citealt{ful96}) are sensitive to radial-velocity variations.

  We select $\eta = 9$ as the value above which lines are
  rejected. These lines are indicated with a ``V'' in
  Table~\ref{linelist4}. The threshold value for $\eta$ is chosen
  arbitrarily, but is a reproducable and objective means to quantify
  line profile variations. As it is an averaged quantity, the $\eta$
  threshold does not reject lines exhibiting only modest EW variations
  and having relatively large errors on the measured
  EWs (e.g. the \ion{H$\epsilon$}{} at 3970.07~{\AA} shown in
  Fig.~\ref{ew_var4}). On the other hand, it does reject some lines
  that have highly accurately determined EWs and that exhibit hardly
  any EW variations with orbital phase. To test the impact of the
  chosen value for the $\eta$ threshold on the obtained value of the
  radial-velocity amplitude, we have evaluated the results for
  different values of $\eta$. Including all lines for which $\eta < 8$
  then $K_{\rm opt}$ is $20.8 \pm 1.5$, $35.7 \pm 1.8$, and $27.5 \pm
  2.3$~\kms\ for \object{SMC~X$-$1}, \object{LMC~X$-$4}, and
  \object{Cen~X$-$3}, respectively. Similarly, we obtain $20.0 \pm
  1.2$, $34.2 \pm 1.3$, and $26.2 \pm 2.7$, respectively, if we select
  the lines with $\eta < 13$. As expected, we obtain a lower value of
  $K_{\rm opt}$ for a higher value of $\eta$, and vice versa, while
  the error on the result increases when including lines that show
  more intrinsic variations. Still, the values for $K_{\rm opt}$ agree
  within the error for the applied range in $\eta$.
  
  We now fix the threshold to $\eta = 9$; to ensure that the
  lines with $\eta < 9$ and small variations concentrated around $\phi
  \sim 0.0$ or $\phi \sim 0.5$, are rejected, we mark these in
  Table~\ref{linelist4} listing the orbital phase at which these
  variations are concentrated. Note that most of these lines are
  already marked with a ``V'', i.e. rejected based on the $\eta$
  threshold.

  The full width at half maximum (FWHM) of the line also varies as a
  function of orbital phase $\phi$. This behaviour is especially
  visible in the Balmer series lines of hydrogen, most prominent in
  the stronger lines. Figure~\ref{fwhm_var4} shows that the FWHM
  increases when the system revolves from X-ray eclipse to $\phi \sim
  0.25$, where it reaches a maximum before declining again. At $\phi
  \sim 0.75$ another, though smaller increase in FWHM is detected. The
  helium (and some other metallic) lines also show FWHM variations,
  but less pronounced and not as periodic as observed in the hydrogen
  lines. Furthermore, these FWHM variations are best seen in
  \object{SMC~X$-$1} and \object{LMC~X$-$4}, but are less clear in
  \object{Cen~X$-$3}.

  Apparently, if one would use these lines to determine the
  (projected) stellar rotation velocity, one would arrive at a larger
  value of $v \sin{i}$ when the system is looked upon from a side
  view. This may be due to the elongated shape of the star as it is
  filling its Roche lobe, as evidenced by the observed ellipsoidal
  variations. The dependence on line strength would be explained by
  the fact that the line forming region is further out in the
  atmosphere when the line is stronger.  This would, however, not
  explain the difference in amplitude of this effect observed between
  $\phi \sim 0.25$ and $\phi \sim 0.75$. 

  Differences in spectral appearance when comparing spectra obtained
  at $\phi \sim 0.25$ and $\phi \sim 0.75$ are well known to occur in
  spectroscopic binaries. The ``Struve-Sahade'' effect
  \citep{str37,sah62} is the apparent strengthening of the secondary
  spectrum of a hot binary when the secondary is approaching and the
  corresponding weakening of the lines when it is receding (see
  \citealt{gie97} for an observational overview). The cause of this
  effect may be the presence of a gas stream (bow shock, wind
  collision) trailing the secondary in its orbit
  \citep{sah59,gie97}. Hydrodynamical simulations of
  \object{SMC~X$-$1} by \citet{blo94} indicate that a collision of
  wind material from the shadow wind with material in the X-ray
  ionisation zone is present in the system. Perhaps that this shocked
  material introduces the difference in amplitude observed in the FWHM
  of the strong Balmer lines that are formed in the stellar wind.

  As we do not have a clear explanation for these variations in FWHM,
  we investigate the possible influence of this effect on the
  determination of $K_{\rm opt}$. Note that the stronger Balmer lines
  are already excluded from the radial-velocity analysis on the basis
  of the EW variations. We fix the FWHM on its mean value, its maximum
  and its minimum and refit the lines. It turns out that the
  measurement of the centre of the line profile is not affected much
  by these FWHM variations. The derived radial-velocity amplitude,
  $K_{\rm opt}$, is the same within its errors in all
  cases. Therefore, we decide not to exclude more lines based on a
  FWHM variation criterion (most of the lines showing this effect
  were excluded on other grounds anyway).

\subsection{Radial-velocity amplitude}
\label{radvel_amplitude4}

  Table~\ref{fitparameters4} shows the final selection of lines for
  which good fits to the radial-velocity curve are obtained, resulting
  in a measurement of $K_{\rm opt}$ and $v_{\gamma}$. Since the phase
  shift should be equal for all lines in one system, we refit all
  lines with the phase shift fixed to the weighted average,
  i.e. $\triangle \phi$ is 0.026, -0.003, and 0.065 for
  \object{SMC~X$-$1}, \object{LMC~X$-$4}, and \object{Cen~X$-$3},
  respectively. Fixing these values does not influence the other
  parameters much; they are the same within their errors.
  \begin{figure}[!t]
  \centering
  \includegraphics[width=7.2cm]{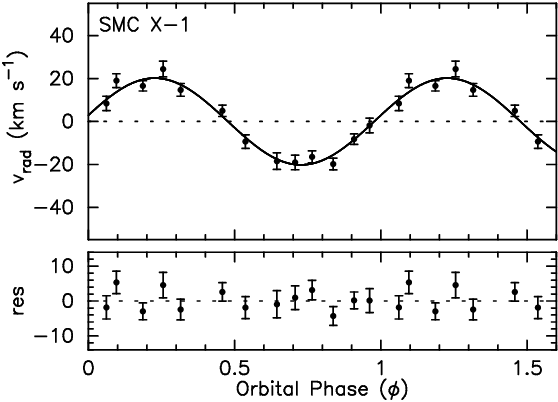}

  \vspace{0.2cm}
  \includegraphics[width=7.2cm]{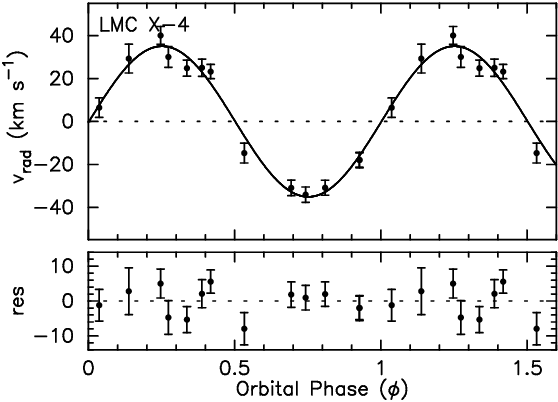}

  \vspace{0.2cm}
  \includegraphics[width=7.2cm]{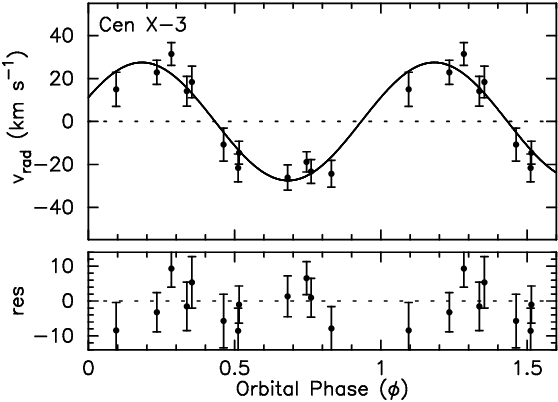}
  \caption{Radial-velocity curves obtained from weighing the radial
  velocities of all selected lines per observation. Some datapoints are
  shown twice to better visualise trends with orbital phase. The error bars
  indicate 1 $\sigma$ errors multiplied by $\chi_r$ to obtain a
  $\chi^2_r = 1$. Note that $v_{\gamma}$ has been set to zero.}
  \label{radvels4}
  \end{figure}

  To avoid sytematic errors while determining the final mean value of
  $K_{\rm opt}$ we shift all determined line centres to equal
  $v_{\gamma}$. The weighted mean values of $v_{\gamma}$ are $191 \pm
  6$~\kms, $306 \pm 10$~\kms\ and $32 \pm 13$~\kms\ for
  \object{SMC~X$-$1}, \object{LMC~X$-$4} and \object{Cen~X$-$3},
  respectively. Then we calculate the weighted mean of the radial
  velocity for each spectrum (see Table~\ref{mean_velocities4}). The
  average dataset we fit in the same way as the individual lines,
  which results in the three radial-velocity curves shown in
  Fig.~\ref{radvels4}. The goodness of the fits with respect to the number of
  degrees of freedom (d.o.f.) are $\chi^2_r/{\rm d.o.f} = 9.8/12$,
  $\chi^2_r/{\rm d.o.f} = 9.1/12$ and $\chi^2_r/{\rm d.o.f} = 7.2/11$
  for \object{SMC~X$-$1}, \object{LMC~X$-$4} and \object{Cen~X$-$3},
  respectively. The error bars indicate 1~$\sigma$ errors multiplied
  by $\chi_r$ to obtain a $\chi^2_r = 1$. The residuals to the fit do
  not show any further evidence for systematic effects with
  orbital phase, as is the case for e.g. \object{Vela~X$-$1}
  \citep{ker95a,bar01}. The final values of $K_{\rm opt}$ are $20.2
  \pm 1.1$~\kms, $35.1 \pm 1.5$~\kms\ and $27.5 \pm 2.3$~\kms\ for
  \object{SMC~X$-$1}, \object{LMC~X$-$4} and \object{Cen~X$-$3},
  respectively. The accuracy of the determination of $K_{\rm opt}$ has
  been significantly improved (by a factor 2--4) compared to previous
  measurements (and comparable to and consistent with \citet{bak05} in
  the case of \object{SMC~X$-$1}). These values are subsequently used
  to calculate the mass of the optical companion and the X-ray source.
  \begin{figure*}[!t]
  \begin{center}
  \includegraphics[width=5.2cm]{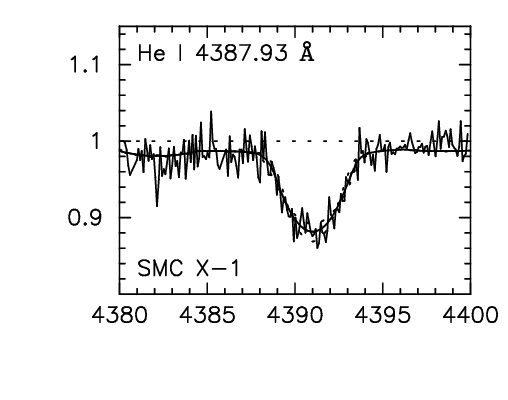}\hspace{-0.6cm}
  \includegraphics[width=5.2cm]{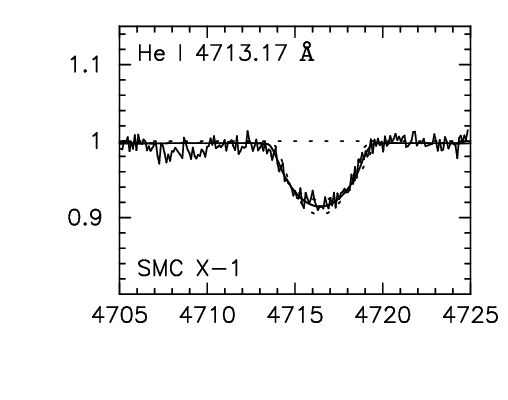}\hspace{-0.6cm}
  \includegraphics[width=5.2cm]{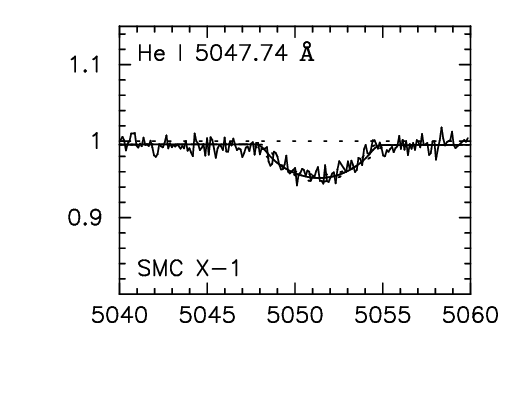}
\\[-0.5cm]
  \includegraphics[width=5.2cm]{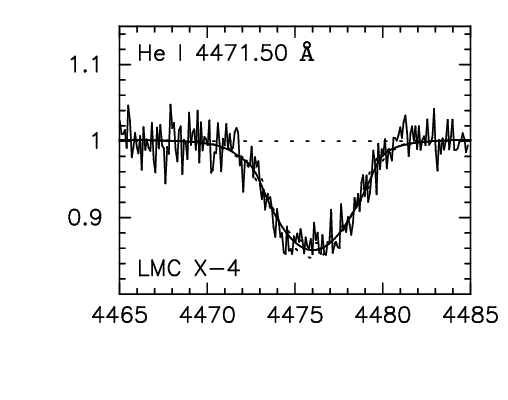}\hspace{-0.6cm}
  \includegraphics[width=5.2cm]{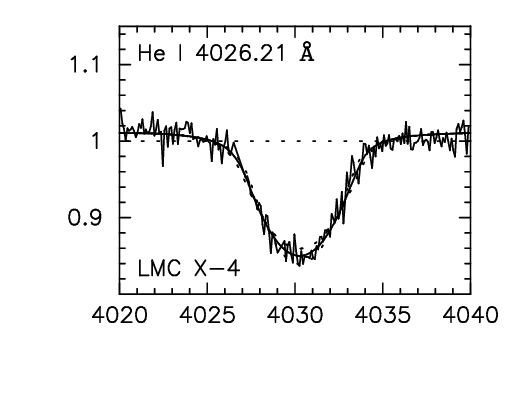}\hspace{-0.6cm}
  \includegraphics[width=5.2cm]{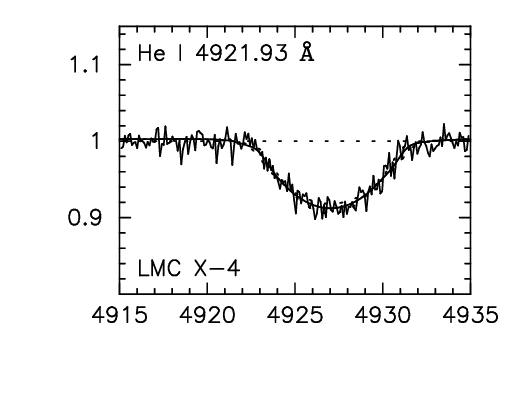}
\\[-0.8cm]
  \includegraphics[width=5.2cm]{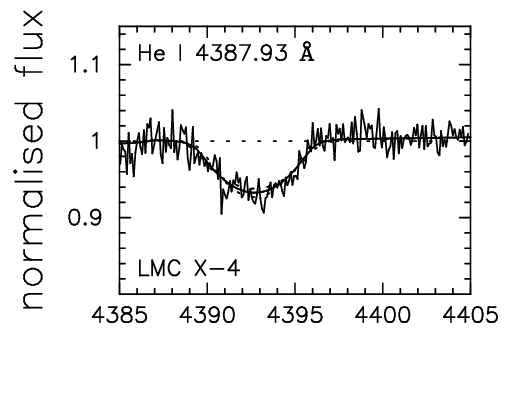}\hspace{-0.6cm}
  \includegraphics[width=5.2cm]{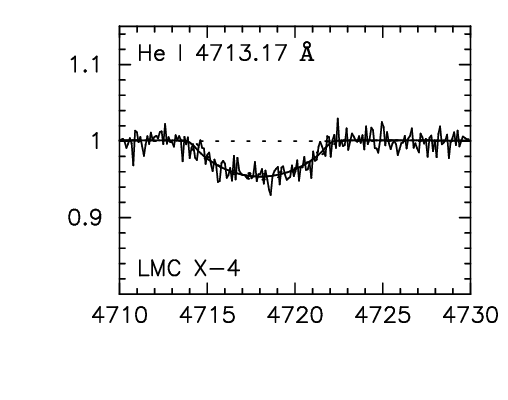}\hspace{-0.6cm}
  \includegraphics[width=5.2cm]{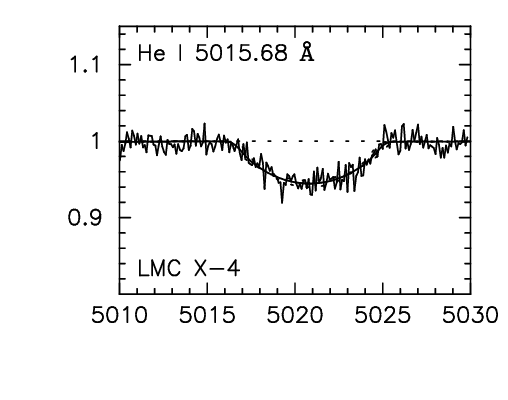}
\\[-0.5cm]
  \includegraphics[width=5.2cm]{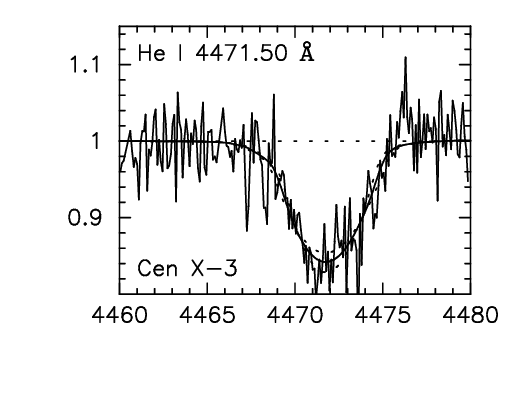}\hspace{-0.6cm}
  \includegraphics[width=5.2cm]{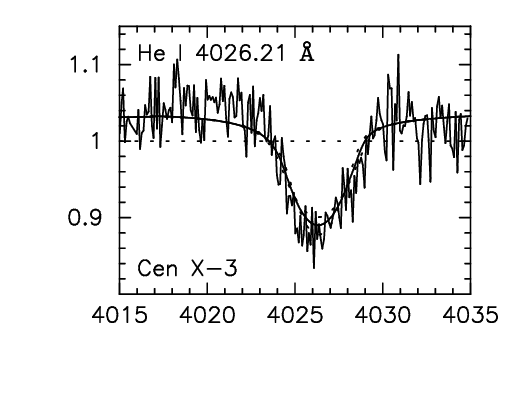}\hspace{-0.6cm}
  \includegraphics[width=5.2cm]{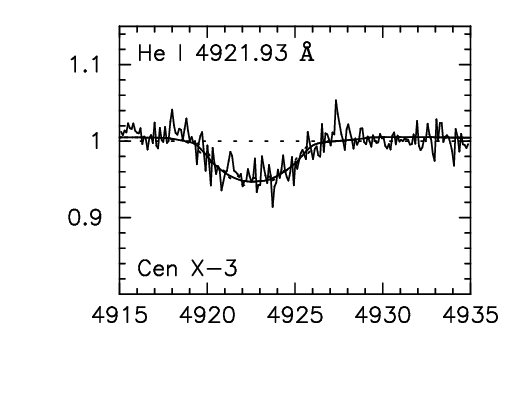}
\\[-0.8cm]
  \includegraphics[width=5.2cm]{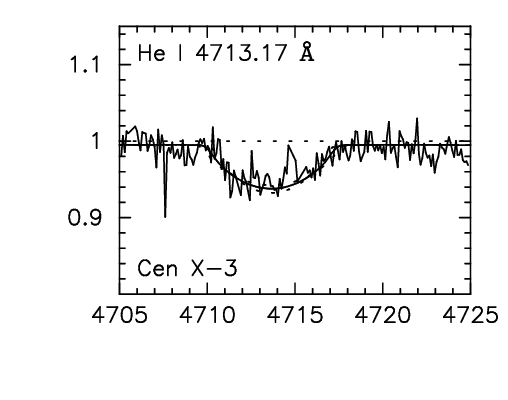}\hspace{-0.6cm}
  \includegraphics[width=5.2cm]{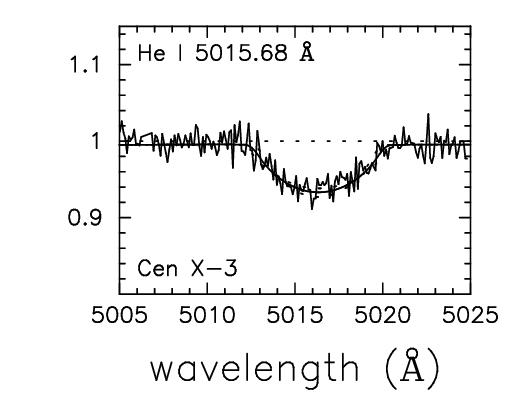}\hspace{4.7cm}
  \caption{Selection of lines for which we measured the rotational
  broadening. Each diagram includes a solid line representing the
  model that we applied; the dotted line represents the
  continuum. To better visualise the model, the spectra are
  smoothed with 7 resolution elements. The top panels show the
  modelled spectral lines of \object{SMC~X$-$1}, the second and third
  row include lines of \object{LMC~X$-$4} and the two bottom rows show
  lines of \object{Cen~X$-$3}.}
  \label{vrots4} 
  \end{center}
  \end{figure*}


\section{The neutron star masses}
\label{nsmasses4}

  In order to measure the mass of the neutron star and its optical
  companion we apply the mass function. For an orbit with eccentricity
  $e$ it can be shown that this is defined as:
  \begin{equation} 
    M_{\rm opt} = \frac{K_{\rm X}^{3}P \left( 1-e^2 \right)^{\frac{3}{2}}} {2\pi G \sin^{3} i} \left( 1+q \right)^{2}
  \end{equation}
  and
  \begin{equation}
    M_{\rm X} = \frac{K_{\rm opt}^{3}P \left( 1-e^2 \right)^{\frac{3}{2}}} {2\pi G \sin^{3} i} \left( 1+\frac{1}{q}\right)^{2},
  \end{equation}
  where $M_{\rm opt}$ and $M_{\rm X}$ are the masses of the optical
  component and the X-ray source, respectively, $K_{\rm opt}$ and
  $K_{\rm X}$ are the semi-amplitude of the radial-velocity curve, $P$
  is the period of the orbit and $i$ is the inclination of the orbital
  plane to the line of sight. The mass ratio $q$ is defined as:
  \begin{equation} 
    q = \frac{M_{\rm X}}{M_{\rm opt}} = \frac{K_{\rm opt}}{K_{\rm X}}
  \end{equation}
 
  The values for $K_{\rm X}$ and $P$ can be obtained very accurately
  from X-ray pulse timing delay measurements
  \citep{woj98,lev00,nag92}. The VLT/UVES spectra provide a value for
  $K_{\rm opt}$. For the determination of the inclination of the
  system we follow the approach of \citet{rap83}, who showed that:
  \begin{equation} 
    \sin i \approx \frac{\sqrt{1 - \beta^{2}\left(\frac{R_{\rm L}}{a}\right)^{2}}} {\cos \theta_{e}} 
  \end{equation} 
  where $R_{\rm L}$ is the Roche-lobe radius of the optical component,
  $\beta$ is the ratio of the radius of the optical component to
  $R_{\rm L}$ (i.e. a Roche-lobe filling factor), $a$ is the
  separation of the centres of mass of the two components, and
  $\theta_{\rm e}$ is the semi-eclipse angle of the compact
  object (see also \citet{jos84} for a review on mass
  determinations in X-ray binaries).
  \begin{figure*}[!t]
  \centering
  \includegraphics[width=4.6cm]{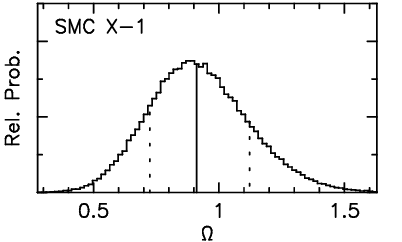}\hspace{-0.6cm}
  \includegraphics[width=4.6cm]{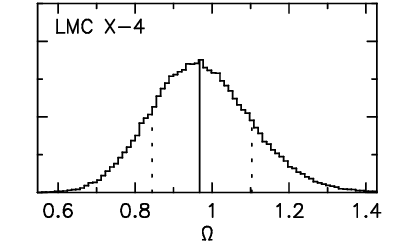}\hspace{-0.6cm}
  \includegraphics[width=4.6cm]{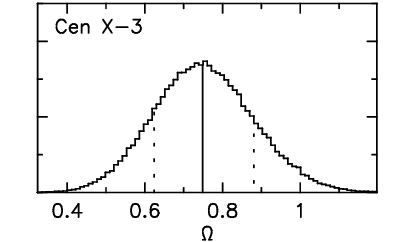}
  \\[0.25cm]
  \includegraphics[width=4.6cm]{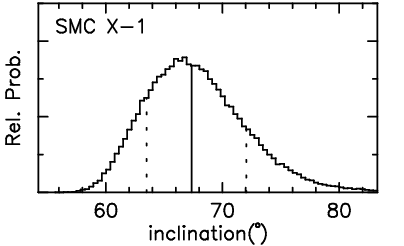}\hspace{-0.6cm}
  \includegraphics[width=4.6cm]{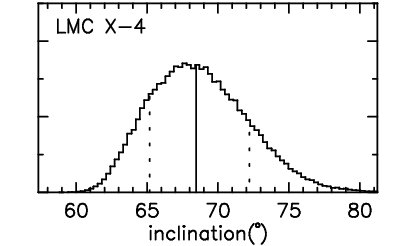}\hspace{-0.6cm}
  \includegraphics[width=4.6cm]{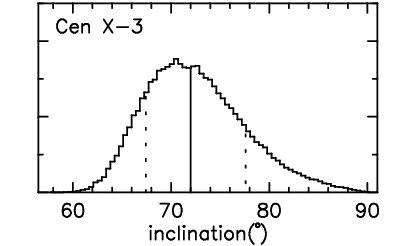}
  \\[0.25cm]
  \includegraphics[width=4.6cm]{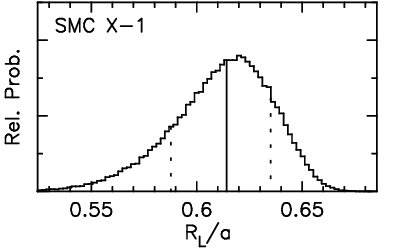}\hspace{-0.6cm}
  \includegraphics[width=4.6cm]{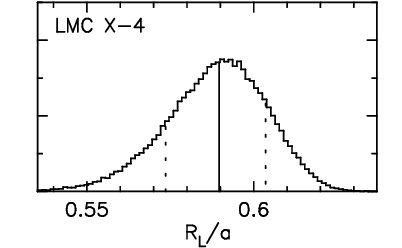}\hspace{-0.6cm}
  \includegraphics[width=4.6cm]{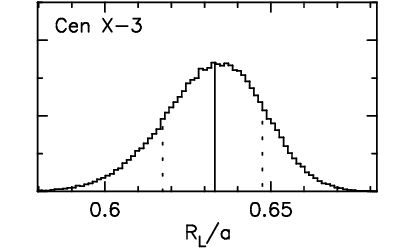}
  \\[0.25cm]
  \includegraphics[width=4.6cm]{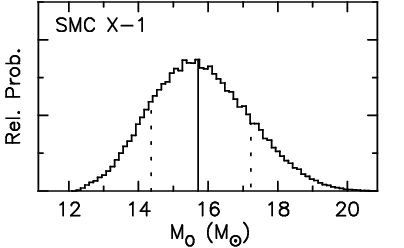}\hspace{-0.6cm}
  \includegraphics[width=4.6cm]{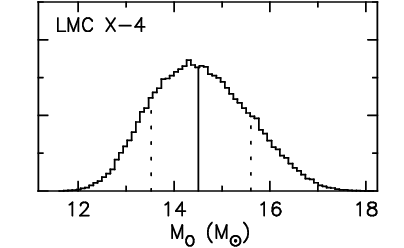}\hspace{-0.6cm}
  \includegraphics[width=4.6cm]{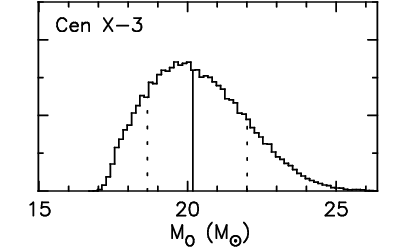}
  \\[0.25cm]
  \includegraphics[width=4.6cm]{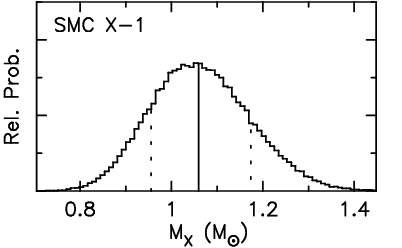}\hspace{-0.6cm}
  \includegraphics[width=4.6cm]{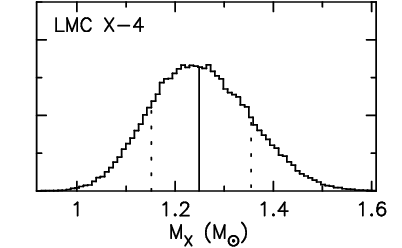}\hspace{-0.6cm}
  \includegraphics[width=4.6cm]{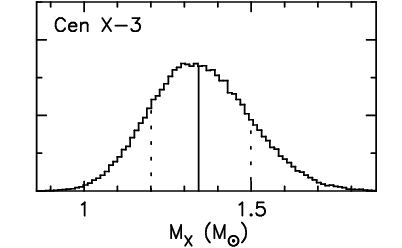}
  \caption{Resulting probability distributions from the Monte-Carlo
  simulations for \object{SMC~X$-$1} (left column), \object{LMC~X$-$4}
  (middle column) and \object{Cen~X$-$3} (rigth column). Shown are the
  ratio of the rotational and orbital frequency ($\Omega$), the
  inclination angle ($i$), the ratio of the Roche-lobe radius and the
  orbital separation ($R_{\rm L} / a$) and the masses of both
  components ($M_{\rm opt}$ \& $M_{\rm X}$) of the three systems.
  The mean value and $1\sigma$ limits are shown with solid and dotted
  lines, respectively.}
  \label{distributions4}
  \end{figure*}

  The ratio of the Roche-lobe radius and the orbital separation can be
  approximated by:
  \begin{equation} 
    \frac{R_{L}}{a} \approx A + B \log q + C \log^{2} q \, .
  \end{equation} 
  The values of the constants $A$, $B$ and $C$ were determined by \citet{rap83} to be:
  \begin{equation} 
    A \approx 0.398 - 0.026 \Omega^{2} + 0.004 \Omega^{3} 
  \end{equation} 

  \begin{equation} 
    B \approx -0.264 + 0.052 \Omega^{2} - 0.015 \Omega^{3} 
  \end{equation} 

  \begin{equation} 
    C \approx -0.023 - 0.005 \Omega^{2},
  \end{equation} 
  where $\Omega$ is the ratio of the rotational frequency of the
  optical companion to the orbital frequency of the system; in case of
  synchronous rotation $\Omega = 1$. However, the timescale at which
  these systems are expected to synchronise is slightly longer than
  the timescale at which the orbit will circularise (for a detailed
  description see e.g. \citealt{hut81}). Therefore, these systems may
  still be in the process of synchronising the optical companion to
  the orbit, whereas their orbits have already become circular. The
  fact that the orbital periods of all three systems are decreasing,
  suggests that the donor stars are rotating slower than synchronous
  and that tidal forces are transferring orbital angular momentum to
  synchronise the system.

  It is possible to determine $\Omega$ by measuring the projected
  rotation velocity $v_{\rm rot} \sin{i}$ of the OB companion using
  the spectra at $\phi = 0.0$. We use the grid of unified stellar
  atmosphere/wind models of early-type supergiants computed by
  \citet{len04} using {\sc cmfgen} \citep{hil98}. First we select the
  lines that correspond to lines included in the model. Subsequently,
  we select the model atmosphere that reproduces the observed line
  spectrum best. The normalised flux of the model is then scaled to
  yield exactly the observed EW. The models that correspond best to
  our selection of lines are named ``AR1Ia'', ``AR1III'' and
  ``O9III-AR2III'' for \object{SMC~X$-$1}, \object{LMC~X$-$4} and
  \object{Cen~X$-$3}, respectively by \citet{len04}. Note that the
  names of these models correspond to a set of model parameters
  describing the model and not to the observational spectral type
  naming convention. These models are subsequently convolved with a
  rotational broadening profile with a limb-darkening coefficient 0.6,
  as described by \citet{gra92}, to determine the value of $v_{\rm
  rot} \sin{i}$. This results in $170 \pm 30$~\kms, $240 \pm 25$~\kms\
  and $200 \pm 40$~\kms\ for \object{SMC~X$-$1}, \object{LMC~X$-$4} and
  \object{Cen~X$-$3}, respectively. The selected lines and their
  corresponding best model are shown in Fig.~\ref{vrots4}. The
  rotational velocities required for a synchronous orbit are
  $185$~\kms, $250$~\kms\ and $255$~\kms\ for \object{SMC~X$-$1},
  \object{LMC~X$-$4} and \object{Cen~X$-$3}, respectively. We will
  find below that these correspond to rotation rates consistent with,
  though perhaps slightly slower than, a synchronous orbit.
  \begin{table*}[!t]
  \caption[]{List of final values for the radial velocity amplitude
  ($K_{\rm opt}$) and the results from the output of the Monte-Carlo
  simulations, i.e. the values of the inclination angle ($i$), the
  ratio of the rotational and orbital frequency ($\Omega$), the ratio
  of the Roche-lobe radius and the orbital separation ($R_{\rm L} /
  a$), the radius of the optical companion ($R_{\rm opt}$), the
  orbital separation ($a$), and the masses of both components ($M_{\rm
  opt}$ \& $M_{\rm X}$) of the three systems. All errors are 1
  $\sigma$.}
  \label{pars_mc4}
  \begin{center}
  \begin{tabular}{p{3.0cm}p{2.5cm}p{2.5cm}p{2.5cm}}
  \hline \hline
  & \object{SMC~X$-$1} & \object{LMC~X$-$4} & \object{Cen~X$-$3} \\
  \hline
  &&&\\[-0.3cm]
  $K_{\rm opt}$ (\kms)       & $ 20.2 \pm 1.1 $       & $ 35.1 \pm 1.5 $       & $ 27.5 \pm 2.3 $ \\
  $i$ ($\deg$)               & $67^{+5}_{-4}$         & $68^{+4}_{-3}$         & $72^{+6}_{-5}$ \\ 
  $\Omega$                   & $0.91^{+0.21}_{-0.19}$ & $0.97^{+0.14}_{-0.12}$ & $0.75 \pm 0.13$ \\
  $R_{\rm L} / a$            & $0.61^{+0.02}_{-0.03}$ & $0.59^{+0.01}_{-0.02}$ & $0.63^{+0.01}_{-0.02}$ \\
  $R_{\rm opt}$ (R$_{\sun}$) & $16.4^{+1.0}_{-1.1}$   & $ 7.8^{+0.3}_{-0.4}$   & $12.1 \pm 0.5$ \\
  $a$ (R$_{\sun}$)           & $26.6 \pm 0.8$         & $13.3 \pm 0.3$         & $19.1^{+0.6}_{-0.5}$ \\
  $M_{\rm opt}$ (M$_{\sun}$) & $15.7^{+1.5}_{-1.4}$   & $14.5^{+1.1}_{-1.0}$   & $20.2^{+1.8}_{-1.5}$ \\
  $M_{\rm X}$ (M$_{\sun}$)   & $1.06^{+0.11}_{-0.10}$ & $1.25^{+0.11}_{-0.10}$ & $1.34^{+0.16}_{-0.14}$ \\
  \hline
  \end{tabular}
  \end{center}
  \end{table*}

  Since for Roche-lobe overflow systems $\beta \gtrsim 0.9$
  \citep{avn75}, we follow the approach of \citet{rap83} and adopt that
  $\beta$ is in the range $0.9-1.0$. Thus, given a set of $K_{\rm
  opt}$, $P$, $a_{\rm X} \sin{i}$ and $\theta_{\rm e}$, we can
  determine by means of Monte-Carlo simulations a 1$\sigma$ confidence
  range for the values of $R_{\rm L} / a$, $i$, $M_{\rm opt}$ and
  $M_{\rm X}$ (see \citealt{rap83,ker95b}). 

  Since in these systems soft X-rays are absorbed by the extended
  stellar wind of the optical companion, the eclipse lasts longer at
  low energies (up to $\sim 5$~keV), depending on the density structure
  of the stellar wind of the optical companion (e.g. for
  \object{4U~1700$-$37} the eclipse at energies up to $\sim 2$~keV
  lasts almost twice as long as at $\sim 6$~keV; see
  \citealt{hab94}; \citealt{mee05}). Therefore, we prefer $\theta_{\rm e}$
  determinations obtained from X-ray observations at high energies
  (see Table~\ref{literature_values4}).  For \object{Cen~X$-$3} these
  measurements are available. An accurate modelling of the X-ray
  light curve of multiple observations obtained with SAS-3/XTCA in the
  energy range $7.9-20$~keV by \citet{cla88} results in a value of
  $\theta_{\rm e} = 32.9\degr \pm 1.4\degr$. They do not list an error
  on their value, so we use the standard deviation in their
  $\theta_{\rm e}$ distribution. For \object{SMC~X$-$1} and
  \object{LMC~X$-$4} no detailed modelling has been performed and
  $\theta_{\rm e}$ measurements are mainly reported for older X-ray
  missions.  We use the range $26\degr-30.5\degr$ for
  \object{SMC~X$-$1} based on observations of \citet{pri76}
  ($28.2\degr \pm 0.9\degr$, $2-6$~keV, SAS-3), \citet{bon81}
  ($29.9\degr \pm 0.2\degr$, $2-12$~keV, COS-B) and \citet{sch72a}
  ($29.1\degr \pm 2.8\degr$, $2-6$~keV, \textit{Uhuru}). For
  \object{LMC~X$-$4} we adopt the range $25\degr-29\degr$ based on
  observations of \citet{li78} ($29.0\degr \pm 2.5\degr$, $6-12$~keV,
  SAS-3), \citet{whi78} ($26.2\degr \pm 1.1\degr$, $2-16$~keV, ARIEL
  V) and \citet{pie85} ($27.1\degr \pm 1.0\degr$, $2-7$~keV, EXOSAT).

  Using the defined input distributions for the Monte-Carlo
  simulations, we can determine the distributions for $i$, $\Omega$,
  $R_{\rm L} / a$, $R_{\rm opt}$, $a$, $M_{\rm opt}$ and $M_{\rm
  X}$. All the results are listed in Table~\ref{pars_mc4} and some of
  the corresponding distributions are shown in
  Fig~\ref{distributions4}. The masses of the neutron stars become:
  $M_{\rm X} = 1.06^{+0.11}_{-0.10}$~M$_{\sun}$ for
  \object{SMC~X$-$1}, $M_{\rm X} = 1.25^{+0.11}_{-0.10}$~M$_{\sun}$
  for \object{LMC~X$-$4} and $M_{\rm X} =
  1.34^{+0.16}_{-0.14}$~M$_{\sun}$ for \object{Cen~X$-$3} at a
  $1\sigma$ confidence level. Compared to the mass determinations
  listed in Sect.~\ref{hmxbs4} the error on the neutron star mass is
  reduced by at least a factor two.

  The masses and radii of the OB~companion stars all lie in the range
  of 15--20~M$_{\sun}$ and 8--16~R$_{\sun}$, respectively, in line
  with previous values \citep{ker95b}. \citet{con78} and \citet{kap01}
  show that for these stars a higher mass is expected based on their
  spectral classification and conclude that these stars are
  undermassive for their luminosity. This may be due to the phase of
  mass transfer prior to the supernova forming the neutron star in the
  system. A detailed modelling of the optical spectra of the OB
  companions will be presented in a forthcoming paper.
  \begin{figure*}[!t]
  \begin{center}
  \includegraphics[width=11.8cm]{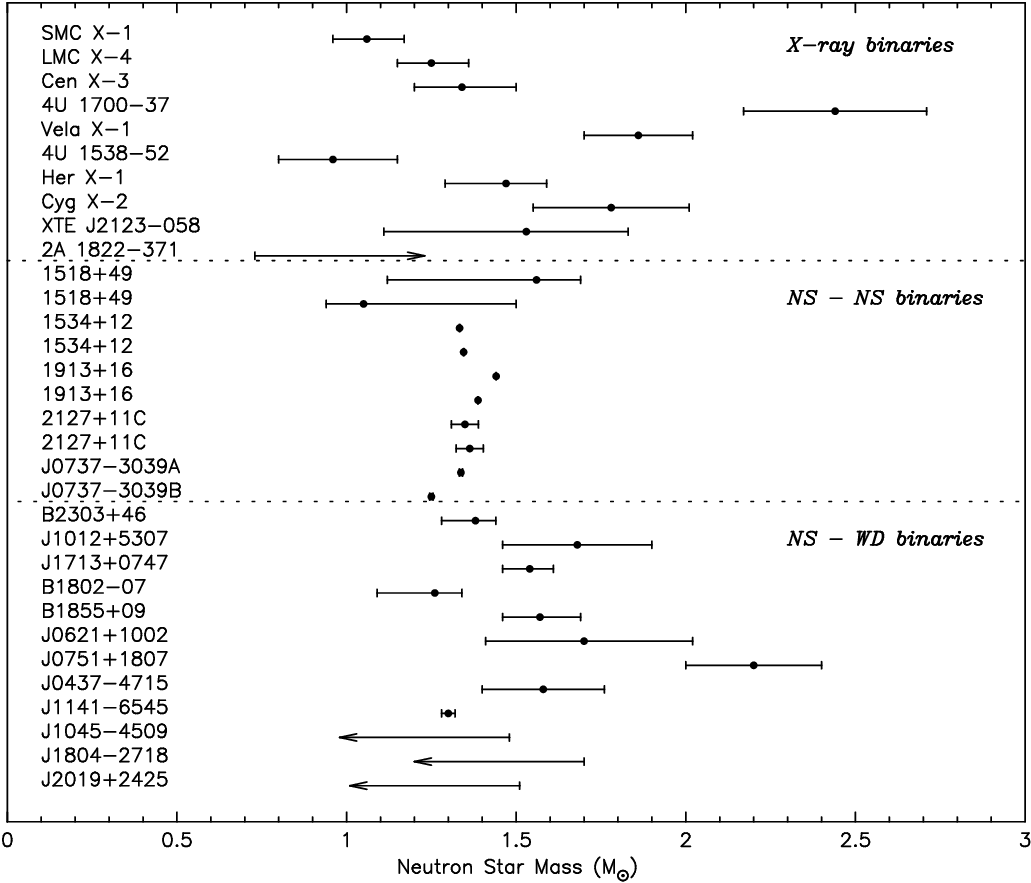}
  \caption{Neutron Star (NS) masses in X-ray binaries, NS$-$NS
  binaries and NS$-$White Dwarf (WD) binaries obtained from
  \citet{sta04} and references therein. The indicated masses of
  \object{SMC~X$-$1}, \object{LMC~X$-$4} and \object{Cen~X$-$3} are
  obtained from this study. The error bars correspond to 1 $\sigma$
  errors. This plot clearly suggests that neutron stars do not all
  have the same ``canonical'' mass. Note that although
  \object{4U~1700$-$37} most probably is a neutron star, it could be a
  black hole.}
  \label{nsmasses_plot4}
  \end{center}
  \end{figure*}


\section{The neutron star mass distribution}
\label{conclusions4}

  \citet{abu04} present a new method to derive the radial velocity
  curves for HMXB systems in which the optical component is deformed
  due to the (partial) filling of its Roche-lobe. They show that these
  systems are affected by gravitational darkening and by X-ray heating
  of the surface of the optical component. This can result in an
  underestimate of the radial velocity amplitude of the optical
  component and therefore in an underestimate of the mass of the
  neutron star. Since this will mostly affect systems hosting a bright
  X-ray source, Roche-lobe overflow systems will suffer most from this
  effect, i.e. the systems discussed in this paper. With our VLT/UVES
  observations it is possible for the first time to accurately
  determine the radial velocity amplitude of each absorption line
  separately. We showed that for lines that vary in EW the radial
  velocity is indeed underestimated by several \kms, consistent with
  the predictions of \citet{abu04}. By applying our carefully chosen
  selection criteria, we anticipate for these effects.

  We conclude that with our observations we have significantly
  improved the accuracy of the determination of the radial-velocity
  amplitude, and subsequently the determination of the neutron star
  mass in these three systems. Whereas some HMXB systems have shown to
  host a neutron star with a mass significantly higher than
  1.4~M$_{\sun}$, as is the case for Vela~X$-$1 and possibly
  4U~1700$-$37, the mass of \object{SMC~X$-$1} is low,
  $1.06^{+0.11}_{-0.10}$~M$_{\sun}$. The masses of \object{LMC~X$-$4}
  and \object{Cen~X$-$3} are $1.25^{+0.11}_{-0.10}$ and
  $1.34^{+0.16}_{-0.14}$~M$_{\sun}$, respectively. The mass of
  \object{SMC~X$-$1} is just above the minimum neutron star mass of
  $\sim 1$~M$_{\sun}$ and significantly different from the mass of the
  neutron star in Vela~X$-$1. We conclude that the neutron stars in
  HMXBs have different masses, i.e. they do not all have the same
  ``canonical'' mass. We illustrate our new mass derivations in
  Fig.~\ref{nsmasses_plot4}, as part of the neutron star masses
  reported by \citet{sta04} and references therein for neutron stars
  in different types of systems.


\section{Discussion}
\label{discussion4}


  It remains to be explained why the mass of \object{SMC~X$-$1} is
  well below 1.28~M$_{\sun}$. The low mass may be the result of a
  different formation scenario, i.e. the electron-capture collapse of
  a degenerate $\ion{O}{}$-$\ion{Ne}{}$-$\ion{Mg}{}$
  core. \citet{heu04} argues that the generally low masses of neutron
  stars measured in binary radio pulsar systems may be due to a
  selection effect, as follows. \citet{pfa02} noticed that there are
  two classes among the wide Be/X-ray binaries: (1) a substantial
  group with low orbital eccentricities, which indicates that their
  neutron stars received hardly any velocity kick in their formation
  events, and (2) a group with high orbital eccentricities, in which
  the neutron stars must have received a kick velocity of several
  hundreds of \kms\ in their birth events.

  It was subsequently noticed \citep{heu04} that the low orbital
  eccentricities of 5 out of the 7 known double neutron stars in the
  galactic disc indicate that the second-born neutron stars in these
  systems received hardly any kick velocity during their birth events
  and thus appear to belong to the same low-kick class of neutron
  stars as the ones in the low-eccentricity Be/X-ray binaries. These
  second-born neutron stars in the low-eccentricity double neutron
  star systems all appear to have low masses, in the range
  $1.18-1.36$~M$_{\sun}$. This fits excellently with neutron-star
  formation by the electron-capture collapse of a degenerate
  $\ion{O}{}$-$\ion{Ne}{}$-$\ion{Mg}{}$ core, which is expected to
  form at the end of the evolution of stars that originated in the
  main-sequence mass-range 8 to about $13 \pm 1$~M$_{\sun}$
  \citep{miy80,pod05,kit05}. Stars with larger masses develop at the end of
  their lives a collapsing iron core, surrounded by convective shells
  with $\ion{O}{}$- and $\ion{Si}{}$-burning. The violent convection
  in these shells may create large density inhomogeneities in the
  layers surrounding the proto-neutron-star formed by the collapsing
  iron core. This may lead to large anisotropies in the neutrino
  transport through these layers, which may cause the neutron star to
  be imparted with a "kick" velocity of some 500~\kms
  \citep{bur96,sch04}. Indeed, young single radio pulsars have large
  space velocities \citep{gun70} and their velocity distribution is
  very well represented by a Maxwellian with a characteristic mean
  velocity of about 400~\kms\ \citep{hob05}.

  In the light of these findings the low mass of the neutron star in
  \object{SMC~X$-$1} would be consistent with its formation by
  electron-capture collapse in a degenerate
  $\ion{O}{}$-$\ion{Ne}{}$-$\ion{Mg}{}$ core. This would imply a
  main-sequence progenitor mass $\lesssim 14$~M$_{\sun}$. Presently the
  companion of \object{SMC~X$-$1} has a mass of about
  $16$~M$_{\sun}$. Allowing for some mass loss by stellar wind, its
  mass just after the mass transfer and the formation of the neutron
  star would have been about $18$~M$_{\sun}$.

  With an explosive mass loss during the formation of the neutron star
  of about $1$~M$_{\sun}$ and a few solar masses stellar wind mass
  loss from the neutron-star progenitor, the initial system must have
  had a mass $\gtrsim 22$~M$_{\sun}$ (including the neutron-star
  mass). Thus, a progenitor system of $13$~M$_{\sun}$+$9$~M$_{\sun}$
  (or $14$~M$_{\sun}$ + $8$~M$_{\sun}$) would be consistent with the
  present system configuration. A potential problem with such a
  configuration is that conservation of mass and orbital angular
  momentum during mass transfer would lead to a fairly wide
  presupernova system, such that the present orbital period of $\sim
  3.9$~days would be hard to understand, unless a large amount of
  orbital angular momentum has been lost with relatively little mass
  (at most a few solar masses) from the system. We thus conclude that
  the low mass of the neutron star in \object{SMC~X$-$1} is consistent
  with formation by electron-capture collapse, provided that a
  relatively large amount of orbital angular momentum was lost from
  the system during the first phase of mass transfer.


\begin{acknowledgements}
  AvdM is supported by the Nederlandse Onderzoekschool voor Astronomie
  (NOVA). We would like to thank Rohied Mokiem for helping us with the
  grid of unified stellar atmosphere/wind models and Godelieve
  Hammerschlag-Hensberge for constructive discussions. The ESO Paranal
  staff is acknowledged for carrying out the sevice mode VLT/UVES
  observations. We are grateful to the anonymous referee for his/her
  constructive comments that helped to improve the quality of the paper.
\end{acknowledgements}

\bibliography{../../references}   

\begin{thebibliography}{96}
\expandafter\ifx\csname natexlab\endcsname\relax\def\natexlab#1{#1}\fi

\bibitem[{{Abubekerov} {et~al.}(2004){Abubekerov}, {Antokhina}, \&
  {Cherepashchuk}}]{abu04}
{Abubekerov}, M.~K., {Antokhina}, E.~A., \& {Cherepashchuk}, A.~M. 2004,
  Astronomy Reports, 48, 89

\bibitem[{{Ash} {et~al.}(1999){Ash}, {Reynolds}, {Roche}, {Norton}, {Still}, \&
  {Morales-Rueda}}]{ash99}
{Ash}, T.~D.~C., {Reynolds}, A.~P., {Roche}, P., {et~al.} 1999, \mnras, 307,
  357

\bibitem[{{Aslanov} \& {Cherepashchuk}(1982)}]{asl82}
{Aslanov}, A.~A. \& {Cherepashchuk}, A.~M. 1982, \azh, 59, 290

\bibitem[{{Avni} \& {Bahcall}(1975)}]{avn75}
{Avni}, Y. \& {Bahcall}, J.~N. 1975, \apj, 197, 675

\bibitem[{{Barziv} {et~al.}(2001){Barziv}, {Kaper}, {Van Kerkwijk}, {Telting},
  \& {Van Paradijs}}]{bar01}
{Barziv}, O., {Kaper}, L., {Van Kerkwijk}, M.~H., {Telting}, J.~H., \& {Van
  Paradijs}, J. 2001, \aap, 377, 925

\bibitem[{{Bildsten} {et~al.}(1997){Bildsten}, {Chakrabarty}, {Chiu}, {Finger},
  {Koh}, {Nelson}, {Prince}, {Rubin}, {Scott}, {Stollberg}, {Vaughan},
  {Wilson}, \& {Wilson}}]{bil97}
{Bildsten}, L., {Chakrabarty}, D., {Chiu}, J., {et~al.} 1997, \apjs, 113, 367

\bibitem[{{Blondin}(1994)}]{blo94}
{Blondin}, J.~M. 1994, \apj, 435, 756

\bibitem[{{Bonnet-Bidaud} \& {Van der Klis}(1981)}]{bon81}
{Bonnet-Bidaud}, J.~M. \& {Van der Klis}, M. 1981, \aap, 97, 134

\bibitem[{{Brown} \& {Bethe}(1994)}]{bro94}
{Brown}, G.~E. \& {Bethe}, H.~A. 1994, \apj, 423, 659

\bibitem[{{Burrows}(2000)}]{burr00}
{Burrows}, A. 2000, \nat, 403, 727

\bibitem[{{Burrows} \& {Hayes}(1996)}]{bur96}
{Burrows}, A. \& {Hayes}, J. 1996, Physical Review Letters, 76, 352

\bibitem[{{Chevalier} \& {Ilovaisky}(1977)}]{che77}
{Chevalier}, C. \& {Ilovaisky}, S.~A. 1977, \aap, 59, L9

\bibitem[{{Chodil} {et~al.}(1967){Chodil}, {Mark}, {Rodrigues}, {Seward},
  {Swift}, {Hiltner}, {Wallerstein}, \& {Mannery}}]{cho67}
{Chodil}, G., {Mark}, H., {Rodrigues}, R., {et~al.} 1967, Physical Review
  Letters, 19, 681

\bibitem[{{Clark} {et~al.}(1988){Clark}, {Minato}, \& {Mi}}]{cla88}
{Clark}, G.~W., {Minato}, J.~R., \& {Mi}, G. 1988, \apj, 324, 974

\bibitem[{{Clarkson} {et~al.}(2003){Clarkson}, {Charles}, {Coe}, {Laycock},
  {Tout}, \& {Wilson}}]{cla03}
{Clarkson}, W.~I., {Charles}, P.~A., {Coe}, M.~J., {et~al.} 2003, \mnras, 339,
  447

\bibitem[{{Conti}(1978)}]{con78}
{Conti}, P.~S. 1978, \aap, 63, 225

\bibitem[{{Conti} \& {Alschuler}(1971)}]{con71}
{Conti}, P.~S. \& {Alschuler}, W.~R. 1971, \apj, 170, 325

\bibitem[{{Crampton} {et~al.}(1985){Crampton}, {Hutchings}, \&
  {Cowley}}]{cra85}
{Crampton}, D., {Hutchings}, J.~B., \& {Cowley}, A.~P. 1985, \apj, 299, 839

\bibitem[{{Dekker} {et~al.}(2000){Dekker}, {D'Odorico}, {Kaufer}, {Delabre}, \&
  {Kotzlowski}}]{dek00}
{Dekker}, H., {D'Odorico}, S., {Kaufer}, A., {Delabre}, B., \& {Kotzlowski}, H.
  2000, in Proc. SPIE Vol. 4008, p. 534-545, Optical and IR Telescope
  Instrumentation and Detectors, Masanori Iye; Alan F. Moorwood; Eds., 534--545

\bibitem[{{Fullerton} {et~al.}(1996){Fullerton}, {Gies}, \& {Bolton}}]{ful96}
{Fullerton}, A.~W., {Gies}, D.~R., \& {Bolton}, C.~T. 1996, \apjs, 103, 475

\bibitem[{{Giacconi} {et~al.}(1971){Giacconi}, {Gursky}, {Kellogg}, {Schreier},
  \& {Tananbaum}}]{gia71}
{Giacconi}, R., {Gursky}, H., {Kellogg}, E., {Schreier}, E., \& {Tananbaum}, H.
  1971, \apjl, 167, L67+

\bibitem[{{Giacconi} {et~al.}(1972){Giacconi}, {Murray}, {Gursky}, {Kellogg},
  {Schreier}, \& {Tananbaum}}]{gia72}
{Giacconi}, R., {Murray}, S., {Gursky}, H., {et~al.} 1972, \apj, 178, 281

\bibitem[{{Gies} {et~al.}(1997){Gies}, {Bagnuolo}, \& {Penny}}]{gie97}
{Gies}, D.~R., {Bagnuolo}, W.~G., \& {Penny}, L.~R. 1997, \apj, 479, 408

\bibitem[{{Gray}(1992)}]{gra92}
{Gray}, D.~F. 1992, {The Observation and Analysis of Stellar Photospheres} (The
  Observation and Analysis of Stellar Photospheres, by David F.~Gray,
  pp.~470.~ISBN 0521408687.~Cambridge, UK: Cambridge University Press, June
  1992.)

\bibitem[{{Gunn} \& {Ostriker}(1970)}]{gun70}
{Gunn}, J.~E. \& {Ostriker}, J.~P. 1970, \apj, 160, 979

\bibitem[{{Haberl} {et~al.}(1994){Haberl}, {Aoki}, \& {Mavromatakis}}]{hab94}
{Haberl}, F., {Aoki}, T., \& {Mavromatakis}, F. 1994, \aap, 288, 796

\bibitem[{{Haensel} {et~al.}(2002){Haensel}, {Zdunik}, \& {Douchin}}]{hae02}
{Haensel}, P., {Zdunik}, J.~L., \& {Douchin}, F. 2002, \aap, 385, 301

\bibitem[{{Heemskerk} \& {Van Paradijs}(1989)}]{hee89}
{Heemskerk}, M.~H.~M. \& {Van Paradijs}, J. 1989, \aap, 223, 154

\bibitem[{{Hilditch} {et~al.}(2005){Hilditch}, {Howarth}, \& {Harries}}]{hil05}
{Hilditch}, R.~W., {Howarth}, I.~D., \& {Harries}, T.~J. 2005, \mnras, 357, 304

\bibitem[{{Hillier} \& {Miller}(1998)}]{hil98}
{Hillier}, D.~J. \& {Miller}, D.~L. 1998, \apj, 496, 407

\bibitem[{{Hiltner} {et~al.}(1972){Hiltner}, {Werner}, \& {Osmer}}]{hil72}
{Hiltner}, W.~A., {Werner}, J., \& {Osmer}, P. 1972, \apjl, 175, L19+

\bibitem[{{Hobbs} {et~al.}(2005){Hobbs}, {Lorimer}, {Lyne}, \&
  {Kramer}}]{hob05}
{Hobbs}, G., {Lorimer}, D.~R., {Lyne}, A.~G., \& {Kramer}, M. 2005, \mnras,
  360, 974

\bibitem[{{Hut}(1981)}]{hut81}
{Hut}, P. 1981, \aap, 99, 126

\bibitem[{{Hutchings} {et~al.}(1979){Hutchings}, {Cowley}, {Crampton}, {Van
  Paradus}, \& {White}}]{hut79}
{Hutchings}, J.~B., {Cowley}, A.~P., {Crampton}, D., {Van Paradus}, J., \&
  {White}, N.~E. 1979, \apj, 229, 1079

\bibitem[{{Hutchings} {et~al.}(1978){Hutchings}, {Crampton}, \&
  {Cowley}}]{hut78}
{Hutchings}, J.~B., {Crampton}, D., \& {Cowley}, A.~P. 1978, \apj, 225, 548

\bibitem[{{Jacoby} {et~al.}(2005){Jacoby}, {Hotan}, {Bailes}, {Ord}, \&
  {Kulkarni}}]{jac05}
{Jacoby}, B.~A., {Hotan}, A., {Bailes}, M., {Ord}, S., \& {Kulkarni}, S.~R.
  2005, \apjl, 629, L113

\bibitem[{{Jones} {et~al.}(1973){Jones}, {Forman}, {Tananbaum}, {Schreier},
  {Gursky}, {Kellogg}, \& {Giacconi}}]{jon73}
{Jones}, C., {Forman}, W., {Tananbaum}, H., {et~al.} 1973, \apjl, 181, L43+

\bibitem[{{Joss} \& {Rappaport}(1984)}]{jos84}
{Joss}, P.~C. \& {Rappaport}, S.~A. 1984, \araa, 22, 537

\bibitem[{{Kaper}(2001)}]{kap01}
{Kaper}, L. 2001, in ASSL Vol. 264: The Influence of Binaries on Stellar
  Population Studies, 125--+

\bibitem[{{Kaper} {et~al.}(1994){Kaper}, {Hammerschlag-Hensberge}, \&
  {Zuiderwijk}}]{kap94}
{Kaper}, L., {Hammerschlag-Hensberge}, G., \& {Zuiderwijk}, E.~J. 1994, \aap,
  289, 846

\bibitem[{{Kaper} \& {Van der Meer}(2005)}]{kap05}
{Kaper}, L. \& {Van der Meer}, A. 2005, ArXiv Astrophysics e-prints

\bibitem[{{Kelley} {et~al.}(1983){Kelley}, {Jernigan}, {Levine}, {Petro}, \&
  {Rappaport}}]{kel83}
{Kelley}, R.~L., {Jernigan}, J.~G., {Levine}, A., {Petro}, L.~D., \&
  {Rappaport}, S. 1983, \apj, 264, 568

\bibitem[{{Kitaura} {et~al.}(2005){Kitaura}, {Janka}, \& {Hillebrandt}}]{kit05}
{Kitaura}, F.~S., {Janka}, H.~., \& {Hillebrandt}, W. 2005, ArXiv Astrophysics
  e-prints

\bibitem[{{Krzeminski}(1974)}]{krz74}
{Krzeminski}, W. 1974, \apjl, 192, L135

\bibitem[{{Lang} {et~al.}(1981){Lang}, {Levine}, {Bautz}, {Hauskins}, {Howe},
  {Primini}, {Lewin}, {Baity}, {Knight}, {Rotschild}, \& {Petterson}}]{lan81}
{Lang}, F.~L., {Levine}, A.~M., {Bautz}, M., {et~al.} 1981, \apjl, 246, L21

\bibitem[{{Lattimer} \& {Prakash}(2004)}]{lat04}
{Lattimer}, J.~M. \& {Prakash}, M. 2004, Science, 304, 536

\bibitem[{{Lennon} {et~al.}(1993){Lennon}, {Dufton}, \& {Fitzsimmons}}]{len93}
{Lennon}, D.~J., {Dufton}, P.~L., \& {Fitzsimmons}, A. 1993, \aaps, 97, 559

\bibitem[{{Lenorzer} {et~al.}(2004){Lenorzer}, {Mokiem}, {de Koter}, \&
  {Puls}}]{len04}
{Lenorzer}, A., {Mokiem}, M.~R., {de Koter}, A., \& {Puls}, J. 2004, \aap, 422,
  275

\bibitem[{{Levine} {et~al.}(1993){Levine}, {Rappaport}, {Deeter}, {Boynton}, \&
  {Nagase}}]{lev93}
{Levine}, A., {Rappaport}, S., {Deeter}, J.~E., {Boynton}, P.~E., \& {Nagase},
  F. 1993, \apj, 410, 328

\bibitem[{{Levine} {et~al.}(2000){Levine}, {Rappaport}, \& {Zojcheski}}]{lev00}
{Levine}, A.~M., {Rappaport}, S.~A., \& {Zojcheski}, G. 2000, \apj, 541, 194

\bibitem[{{Li} {et~al.}(1978){Li}, {Rappaport}, \& {Epstein}}]{li78}
{Li}, F., {Rappaport}, S., \& {Epstein}, A. 1978, \nat, 271, 37

\bibitem[{{Liller}(1973)}]{lil73}
{Liller}, W. 1973, \apjl, 184, L37+

\bibitem[{{Lutovinov} {et~al.}(2005){Lutovinov}, {Revnivtsev}, {Gilfanov},
  {Shtykovskiy}, {Molkov}, \& {Sunyaev}}]{lut05}
{Lutovinov}, A., {Revnivtsev}, M., {Gilfanov}, M., {et~al.} 2005, \aap, 444,
  821

\bibitem[{{Mason} {et~al.}(1976){Mason}, {Branduardi}, \& {Sanford}}]{mas76}
{Mason}, K.~O., {Branduardi}, G., \& {Sanford}, P. 1976, \apjl, 203, L29

\bibitem[{{Miyaji} {et~al.}(1980){Miyaji}, {Nomoto}, {Yokoi}, \&
  {Sugimoto}}]{miy80}
{Miyaji}, S., {Nomoto}, K., {Yokoi}, K., \& {Sugimoto}, D. 1980, \pasj, 32, 303

\bibitem[{{Mokiem} {et~al.}(2005){Mokiem}, {de Koter}, {Puls}, {Herrero},
  {Najarro}, \& {Villamariz}}]{mok05}
{Mokiem}, M.~R., {de Koter}, A., {Puls}, J., {et~al.} 2005, \aap, 441, 711

\bibitem[{{Nagase} {et~al.}(1992){Nagase}, {Corbet}, {Day}, {Inoue},
  {Takeshima}, {Yoshida}, \& {Mihara}}]{nag92}
{Nagase}, F., {Corbet}, R.~H.~D., {Day}, C.~S.~R., {et~al.} 1992, \apj, 396,
  147

\bibitem[{{Negueruela} {et~al.}(2006){Negueruela}, {Smith}, {Reig}, {Chaty}, \&
  {Torrej{\'o}n}}]{neg06}
{Negueruela}, I., {Smith}, D.~M., {Reig}, P., {Chaty}, S., \& {Torrej{\'o}n},
  J.~M. 2006, in ESA Special Publication, Vol. 604, ESA Special Publication,
  ed. A.~{Wilson}, 165--170

\bibitem[{{Nice} {et~al.}(2005){Nice}, {Splaver}, {Stairs}, {Loehmer},
  {Jessner}, {Kramer}, \& {Cordes}}]{nic05}
{Nice}, D., {Splaver}, E., {Stairs}, I., {et~al.} 2005, ArXiv Astrophysics
  e-prints

\bibitem[{{Paul} {et~al.}(2005){Paul}, {Raichur}, \& {Mukherjee}}]{pau05}
{Paul}, B., {Raichur}, H., \& {Mukherjee}, U. 2005, ArXiv Astrophysics e-prints

\bibitem[{{Petro} \& {Hiltner}(1982)}]{pet82}
{Petro}, L.~D. \& {Hiltner}, W.~A. 1982, NASA STI/Recon Technical Report N, 85,
  19905

\bibitem[{{Pfahl} {et~al.}(2002){Pfahl}, {Rappaport}, {Podsiadlowski}, \&
  {Spruit}}]{pfa02}
{Pfahl}, E., {Rappaport}, S., {Podsiadlowski}, P., \& {Spruit}, H. 2002, \apj,
  574, 364

\bibitem[{{Pietsch} {et~al.}(1985){Pietsch}, {Voges}, {Pakull}, \&
  {Staubert}}]{pie85}
{Pietsch}, W., {Voges}, W., {Pakull}, M., \& {Staubert}, R. 1985, Space Science
  Reviews, 40, 371

\bibitem[{{Podsiadlowski} {et~al.}(2005){Podsiadlowski}, {Dewi}, {Lesaffre},
  {Miller}, {Newton}, \& {Stone}}]{pod05}
{Podsiadlowski}, P., {Dewi}, J.~D.~M., {Lesaffre}, P., {et~al.} 2005, \mnras,
  361, 1243

\bibitem[{{Priedhorsky} \& {Terrell}(1983)}]{pri83}
{Priedhorsky}, W.~C. \& {Terrell}, J. 1983, \apj, 273, 709

\bibitem[{{Primini} {et~al.}(1976){Primini}, {Clark}, {Lewin}, {Li}, {Mayer},
  {McClintock}, {Rappaport}, \& {Joss}}]{pri76}
{Primini}, F., {Clark}, G.~W., {Lewin}, W., {et~al.} 1976, \apjl, 210, L71

\bibitem[{{Quaintrell} {et~al.}(2003){Quaintrell}, {Norton}, {Ash}, {Roche},
  {Willems}, {Bedding}, {Baldry}, \& {Fender}}]{qua03}
{Quaintrell}, H., {Norton}, A.~J., {Ash}, T.~D.~C., {et~al.} 2003, \aap, 401,
  313

\bibitem[{{Rappaport} \& {Joss}(1983)}]{rap83}
{Rappaport}, S.~A. \& {Joss}, P.~C. 1983, in Accretion-Driven Stellar X-ray
  Sources, 1--39

\bibitem[{{Reynolds} {et~al.}(1992){Reynolds}, {Bell}, \& {Hilditch}}]{rey92}
{Reynolds}, A.~P., {Bell}, S.~A., \& {Hilditch}, R.~W. 1992, \mnras, 256, 631

\bibitem[{{Reynolds} {et~al.}(1993){Reynolds}, {Hilditch}, {Bell}, \&
  {Hill}}]{rey93}
{Reynolds}, A.~P., {Hilditch}, R.~W., {Bell}, S.~A., \& {Hill}, G. 1993,
  \mnras, 261, 337

\bibitem[{{Reynolds} {et~al.}(1999){Reynolds}, {Owens}, {Kaper}, {Parmar}, \&
  {Segreto}}]{rey99}
{Reynolds}, A.~P., {Owens}, A., {Kaper}, L., {Parmar}, A.~N., \& {Segreto}, A.
  1999, \aap, 349, 873

\bibitem[{{Sahade}(1959)}]{sah59}
{Sahade}, J. 1959, \pasp, 71, 151

\bibitem[{{Sahade}(1962)}]{sah62}
{Sahade}, J. 1962, in Proceedings of a Symposium on Stellar Evolution, held in
  La Plata, November 7-11, 1960, La Plata: National University, Astronomical
  Observatory, 1962, edited by Sahade, J., p.185, 185--+

\bibitem[{{Sanduleak} \& {Philip}(1976)}]{san76}
{Sanduleak}, N. \& {Philip}, A.~G.~D. 1976, \iaucirc, 3023, 1

\bibitem[{{Savonije}(1978)}]{sav78}
{Savonije}, G.~J. 1978, \aap, 62, 317

\bibitem[{{Savonije}(1983)}]{sav83}
{Savonije}, J. 1983, in Accretion-Driven Stellar X-ray Sources, 343--366

\bibitem[{{Scheck} {et~al.}(2004){Scheck}, {Plewa}, {Janka}, {Kifonidis}, \&
  {M{\" u}ller}}]{sch04}
{Scheck}, L., {Plewa}, T., {Janka}, H.-T., {Kifonidis}, K., \& {M{\" u}ller},
  E. 2004, Physical Review Letters, 92, 011103

\bibitem[{{Schreier} {et~al.}(1972{\natexlab{a}}){Schreier}, {Giacconi},
  {Gursky}, {Kellogg}, \& {Tananbaum}}]{sch72a}
{Schreier}, E., {Giacconi}, R., {Gursky}, H., {Kellogg}, E., \& {Tananbaum}, H.
  1972{\natexlab{a}}, \apjl, 178, L71+

\bibitem[{{Schreier} {et~al.}(1972{\natexlab{b}}){Schreier}, {Levinson},
  {Gursky}, {Kellogg}, {Tananbaum}, \& {Giacconi}}]{sch72b}
{Schreier}, E., {Levinson}, R., {Gursky}, H., {et~al.} 1972{\natexlab{b}},
  \apjl, 172, L79+

\bibitem[{{Schrijvers} \& {Telting}(1999)}]{sch99}
{Schrijvers}, C. \& {Telting}, J.~H. 1999, \aap, 342, 453

\bibitem[{{Schrijvers} {et~al.}(1997){Schrijvers}, {Telting}, {Aerts},
  {Ruymaekers}, \& {Henrichs}}]{sch97}
{Schrijvers}, C., {Telting}, J.~H., {Aerts}, C., {Ruymaekers}, E., \&
  {Henrichs}, H.~F. 1997, \aaps, 121, 343

\bibitem[{{Srinivasan}(2001)}]{sri01}
{Srinivasan}, G. 2001, in Black Holes in Binaries and Galactic Nuclei, 45--+

\bibitem[{{Stairs}(2004)}]{sta04}
{Stairs}, I.~H. 2004, Science, 304, 547

\bibitem[{{Struve}(1937)}]{str37}
{Struve}, O. 1937, \apj, 85, 41

\bibitem[{{Thorsett} \& {Chakrabarty}(1999)}]{tho99}
{Thorsett}, S.~E. \& {Chakrabarty}, D. 1999, \apj, 512, 288

\bibitem[{{Timmes} {et~al.}(1996){Timmes}, {Woosley}, \& {Weaver}}]{tim96}
{Timmes}, F.~X., {Woosley}, S.~E., \& {Weaver}, T.~A. 1996, \apj, 457, 834

\bibitem[{{Tjemkes} {et~al.}(1986){Tjemkes}, {Van Paradijs}, \&
  {Zuiderwijk}}]{tje86}
{Tjemkes}, S.~A., {Van Paradijs}, J., \& {Zuiderwijk}, E.~J. 1986, \aap, 154,
  77

\bibitem[{{Val Baker} {et~al.}(2005){Val Baker}, {Norton}, \&
  {Quaintrell}}]{bak05}
{Val Baker}, A.~K.~F., {Norton}, A.~J., \& {Quaintrell}, H. 2005, \aap, 441,
  685

\bibitem[{{Van den Heuvel}(2004)}]{heu04}
{Van den Heuvel}, E.~P.~J. 2004, in ESA SP-552: 5th INTEGRAL Workshop on the
  INTEGRAL Universe, 185--+

\bibitem[{{Van der Meer} {et~al.}(2005){Van der Meer}, {Kaper}, {di Salvo},
  {M{\' e}ndez}, {Van der Klis}, {Barr}, \& {Trams}}]{mee05}
{Van der Meer}, A., {Kaper}, L., {di Salvo}, T., {et~al.} 2005, \aap, 432, 999

\bibitem[{{Van Kerkwijk} {et~al.}(1995{\natexlab{a}}){Van Kerkwijk}, {Van
  Paradijs}, \& {Zuiderwijk}}]{ker95b}
{Van Kerkwijk}, M.~H., {Van Paradijs}, J., \& {Zuiderwijk}, E.~J.
  1995{\natexlab{a}}, \aap, 303, 497

\bibitem[{{Van Kerkwijk} {et~al.}(1995{\natexlab{b}}){Van Kerkwijk}, {Van
  Paradijs}, {Zuiderwijk}, {Hammerschlag-Hensberge}, {Kaper}, \&
  {Sterken}}]{ker95a}
{Van Kerkwijk}, M.~H., {Van Paradijs}, J., {Zuiderwijk}, E.~J., {et~al.}
  1995{\natexlab{b}}, \aap, 303, 483

\bibitem[{{Van Paradijs} {et~al.}(1978){Van Paradijs},
  {Hammerschlag-Hensberge}, \& {Zuiderwijk}}]{par78}
{Van Paradijs}, J.~A., {Hammerschlag-Hensberge}, G., \& {Zuiderwijk}, E.~J.
  1978, \aaps, 31, 189

\bibitem[{{Vidal} {et~al.}(1973){Vidal}, {Wickramsinghe}, \&
  {Peterson}}]{vid73}
{Vidal}, N.~V., {Wickramsinghe}, D.~T., \& {Peterson}, B.~A. 1973, \apjl, 182,
  L77+

\bibitem[{{White}(1978)}]{whi78}
{White}, N.~E. 1978, \nat, 271, 38

\bibitem[{{Wojdowski} {et~al.}(1998){Wojdowski}, {Clark}, {Levine}, {Woo}, \&
  {Zhang}}]{woj98}
{Wojdowski}, P., {Clark}, G.~W., {Levine}, A.~M., {Woo}, J.~W., \& {Zhang},
  S.~N. 1998, \apj, 502, 253

\end{thebibliography}
   
\end{document}